\documentclass[superscriptaddress,notitlepage,longbibliography]{revtex4-1}
\usepackage{caption}
\usepackage{pifont}
\usepackage{listings}
\usepackage{color}
\usepackage{verbatim}
\usepackage{graphicx}
\usepackage{amsmath}
\usepackage[utf8x]{inputenc}
\usepackage[colorlinks=true,urlcolor=blue,citecolor=blue]{hyperref}

\newcommand{\Xv}[1]{\langle #1\rangle}
\definecolor{mygreen}{rgb}{0,0.6,0}
\definecolor{lgray}{rgb}{0.9,0.9,0.9}
\definecolor{mygray}{rgb}{0.5,0.5,0.5}
\definecolor{mymauve}{rgb}{0.58,0,0.82}
\definecolor{darkolivegreen}{rgb}{0.33, 0.42, 0.18}
\definecolor{scala-orange}{rgb}{0.84, 0.33, 0}
\definecolor{scala-blue}{rgb}{0, 0.55, 0.84}
\definecolor{scala-yellow}{rgb}{0.84, 0.67, 0}
\definecolor{scala-aqua}{rgb}{0, 0.68, 0.41}
\definecolor{scala-purple}{rgb}{0.51, 0, 0.92}
\begin{document}
\title{Evaluating probabilistic programming languages for simulating quantum correlations}

\author{Abdul Obeid}
\affiliation{School of Information Systems, Queensland University of Technology, Australia}
\author{Peter D. Bruza}
\affiliation{School of Information Systems, Queensland University of Technology, Australia}
\author{Peter Wittek}
\affiliation{Rotman School of Management, University of Toronto, M5S 3E6 Toronto, Canada}
\affiliation{Creative Destruction Lab, M5S 3E6 Toronto, Canada}
\affiliation{Vector Institute for Artificial Intelligence, M5G 1M1 Toronto, Canada}
\affiliation{Perimeter Institute for Theoretical Physics, N2L 2Y5 Waterloo, Canada}

\begin{abstract}
This article explores how probabilistic programming can be used to simulate quantum correlations in an EPR experimental setting.
Probabilistic programs are based on standard probability which cannot produce quantum correlations.
In order to address this limitation, a hypergraph formalism was programmed which both expresses the measurement contexts of the EPR experimental design as well as associated constraints. Four contemporary open source probabilistic programming frameworks were used to simulate an EPR experiment in order to shed light on their relative effectiveness from both qualitative and quantitative dimensions.
We found that all four probabilistic languages successfully simulated quantum correlations. Detailed analysis revealed that no language was clearly superior across all dimensions, however, the comparison does highlight aspects that can be considered when using probabilistic programs to simulate experiments in quantum physics.
\end{abstract}

\maketitle

\section{Introduction}
Probabilistic models are used in a broad swathe of disciplines ranging from the social and behavioural sciences, biology, the physical and computational sciences, to name but a few. 
At their very core, probabilistic models are defined in terms of random variables, which range over a set of outcomes that are subject to chance.
For example, a measurement on a quantum system is a random variable.
By performing the measurement, we record the outcome as the value of the random variable.
Repeated measurements on the same preparation allow determining the probability of each outcome.
Probabilistic programming offers a convenient way to express probabilistic models by unifying techniques from conventional programming such as modularity, imperative or functional specification, as well as the representation and use of uncertain knowledge. 
A variety of probabilistic programming languages (PPLs) have been proposed (see \cite{gordon:henzinger:2014} for references), which have attracted interest from artificial intelligence, programming languages, cognitive science, and the natural language processing communities \cite{goodman:stuhlmuller:2014}.
However, as far as the authors can tell, there has been little interest in PPLs from the physics research community.
The aim of this article is raise awareness of PPLs to this community by showing how quantum correlations can be simulated by probabilistic programming.

The core purpose of a PPL is to specify a model in terms of random variables and probability distributions \cite{gordon:henzinger:2014}.
As a consequence, a PPL is restricted to computing statistical correlations between variables which are a mathematical consequence of the underlying event space.
Quantum theory, on the the hand, has a different underlying event space.
This in turn allows correlations between variables to emerge that go beyond those governed by standard probability theory.
In particular, local hidden variables are straightforward to represent in a PPL, since they correspond to what classical probabilities can express.
Nonlocal correlations, however, cannot be described by a local hidden variable model~\cite{bell1964epr}. 
The question arises as to how to simulate such correlations using probabilistic programming.
This article addresses this question by using a hypergraph formalism that has recently emerged in quantum information~\cite{acin:2015}. 
The advantage of the hypergraph formalism is that it provides a flexible, abstract representation for rendering into the syntax of a PPL.
In addition, constraints inherent to the experimental design being simulated can be structurally expressed within the hypergraphs.
We will show that by embedding this hypergraph formalism in a PPL, an EPR\footnote{The acronym EPR describes from Einstein, Rosen and Podolky's famous paper which subsequently led to experimental protocols being developed to investigate quantum entanglement~\cite{einstein1935can}.} experiment can be simulated where quantum correlations are produced.
In addition, we provide qualitative and quantitative comparisons between several implementations in contemporary PPLs under an open source license~\footnote{The code is available at \url{https://github.com/askoj/bell-ppls}}.
This opens the door to the possibility of reliably and meaningfully simulating experiments in quantum contextuality by means of probabilistic programs.

\section{Probabilistic Programming and the ERP experiment}
\begin{figure}[h]
\centering
\includegraphics[width=5.25cm]{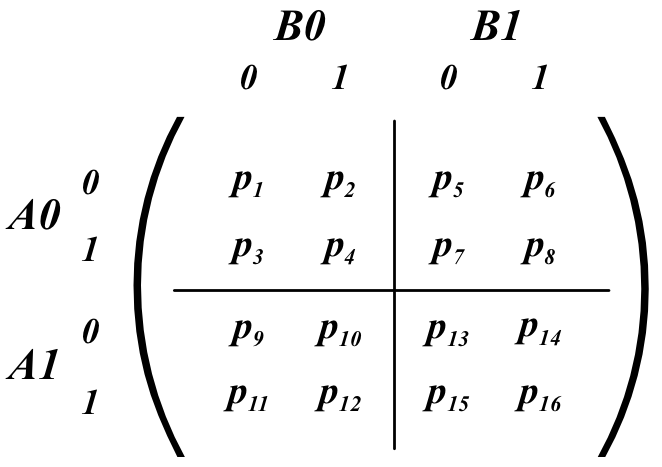}
\centering
\captionsetup{justification=centering,margin=2cm}
\caption {Four pairwise distributions in an EPR experiment}
\label{fig:p16}
\end{figure}

The basis of the EPR experiment is two systems $A$ and $B$ which are represented as bivalent variables ranging of $\{0,1\}$.
Variables $A$ and $B$ are respectively conditioned by bivalent variables $X$ and $Y$, with both ranging over $\{0,1\}$ .
Four experiments are performed by means of joint measurements on $A$ and $B$ depending on the value of the respective conditioning variables.
As a consequence, the experiments produce four pairwise distributions over the four possible outcomes from the joint measurements:
\begin{align*}
p(A,B|X=0,Y=0) \\
p(A,B|X=0,Y=1) \\
p(A,B|X=1,Y=0) \\
p(A,B|X=1,Y=1) 
\end{align*}
In order to simplify the notation, 
variable $A_i$ is distributed as $p(A|X=i), i \in \{0,1\}$.
In a similar way, variables $B_0$ and $B_1$ are introduced.
Therefore, the preceding four pairwise distributions can be represented as the the grid of sixteen probabilities depicted in Fig.~\ref{fig:p16}.
The EPR experiment is subject to constraint know as the ``no-signalling" condition.
No-signalling entails that the marginal probabilities observed in relation to one variable do not vary according to how the other variable is conditioned:
\begin{align}
p_1 + p_2 = p_5 + p_6 \\
p_9 + p_{10} = p_{13} + p_{14} \\
p_1 + p_3 = p_9 + p_{11} \\
p_5 + p_7 = p_{13} + p_{15}
\end{align}
The goal of an EPR experiment is to empirically determine whether quantum particles are entangled.
We will not go into the details of what entanglement is, but rather focus on showing how statistical correlations between variables determine the presence of entanglement. 
Entanglement is determined if any of the following inequalities is violated.
\begin{eqnarray}
|\Xv{A_{0}B_{0}} + \Xv{A_{0}B_{1}} + \Xv{A_{1}B_{0}} - \Xv{A_{1}B_{1}} | &\leq 2   \label{eqn:chsh1}\\
|\Xv{A_{0}B_{0}} + \Xv{A_{0}B_{1}} - \Xv{A_{1}B_{0}} + \Xv{A_{1}B_{1}} | &\leq 2  \label{eqn:chsh2}  \\
|\Xv{A_{0}B_{0}} - \Xv{A_{0}B_{1}} + \Xv{A_{1}B_{0}} + \Xv{A_{1}B_{1}} | &\leq 2  \label{eqn:chsh3}  \\
|-\Xv{A_{0}B_{0}} + \Xv{A_{0}B_{1}} + \Xv{A_{1}B_{0}} + \Xv{A_{0}B_{0}} | &\leq 2   \label{eqn:chsh4} 
\end{eqnarray}
where the correlations are defined as follows:
\begin{align}
\Xv{A_{0}B_{0}} &= (p_1+p_4) - (p_2 + p_3) \\
\Xv{A_{0}B_{1}} &= (p_5+p_8) - (p_6 + p_7) \\
\Xv{A_{1}B_{0}} &= (p_9+p_{12}) - (p_{10} + p_{11}) \\
\Xv{A_{1}B_{1}} &= (p_{13}+p_{16}) - (p_{14} + p_{15}) 
\end{align}
For historical reasons, the set of four inequalities have become known as the Clauser-Horn-Shimony-Holt (CHSH) inequalities \cite{Shimony:Bell}. 
The data is collected from the four experiments defined above by subjecting a large number of pairs $(A,B)$ of quantum particles to joint measurements. More specifically, each such pair is measured in one of the four measurement conditions represented by the grid of probabilities depicted in Fig.~\ref{fig:p16}.

The maximum possible violation of the CHSH inequalities is 4, i.e., three pairs of variables are maximally correlated (=1) and the fourth is maximally anti-correlated (=-1).
However, if the experiment is modelled by a joint probability distribution across the four variables $A_0,A_1,B_0,B_1$, the maximum value that can be computed by any of the inequalities happens to be 2. 
This is why the boundary of violation in the inequalities is 2 as it demarcates the boundary which standard statistical correlations cannot transcend.
This fact presents a challenge for a PPL, which is based on standard probability theory. 
How can a PPL be developed to simulate non-classical quantum correlations?

\section{Design of an EPR Simulation Experiment using PPLs}
\begin{figure}[h]
\captionsetup{justification=centering}
\centering
\includegraphics[width=12cm]{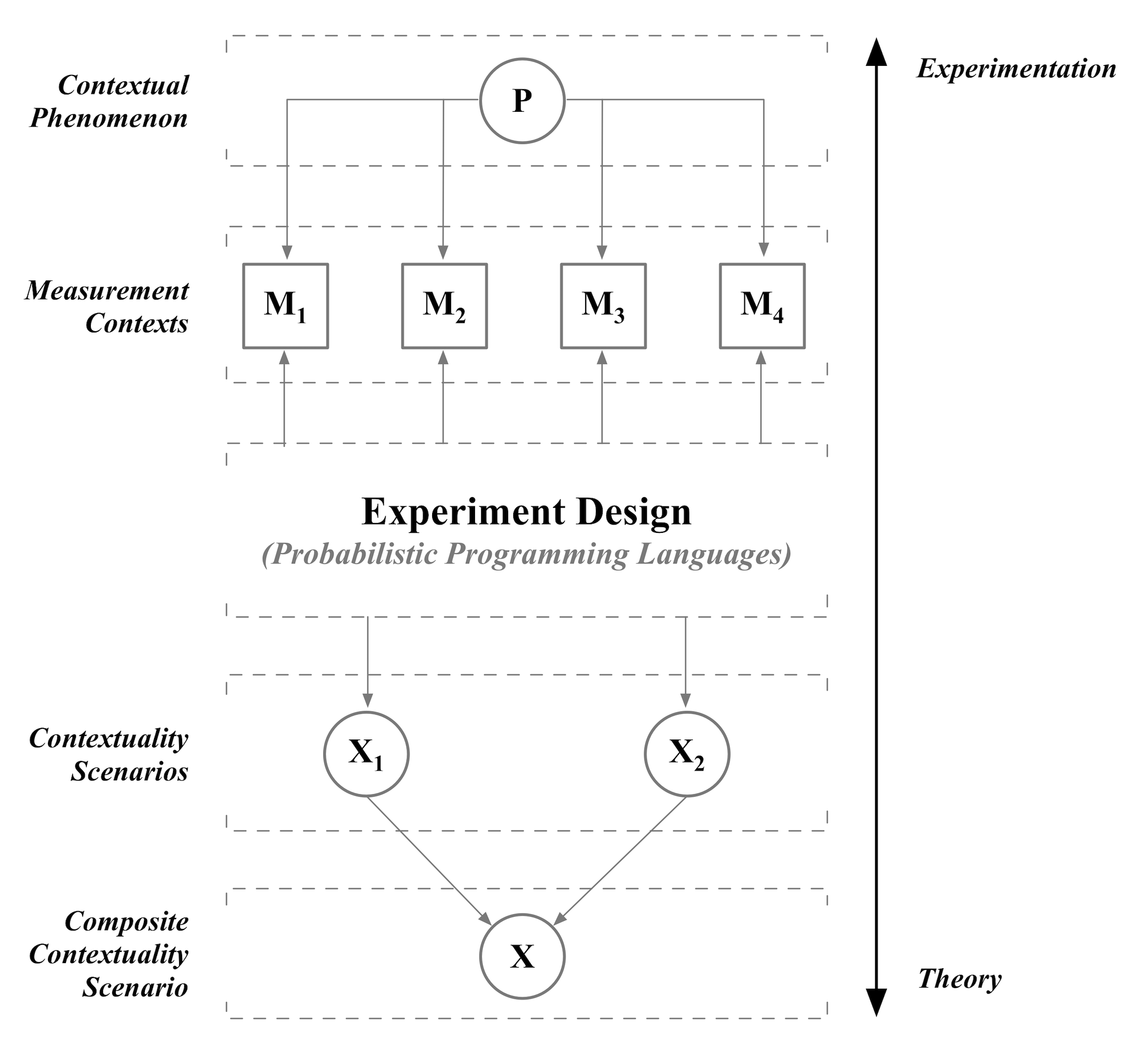}
\caption{Framework for the EPR experiment}\label{fig:PPL-framework}
\end{figure}
Fig.~\ref{fig:PPL-framework} depicts the framework for how 
a PPL can be used to simulate EPR experiments. 
A phenomenon $P$, e.g., entangled quantum particles, is to be studied. 
An experimental design is devised in which $P$ is examined in the four experimental conditions called ``measurement contexts". 
A measurement context $M_i, 1 \leq i \leq 4$ is designed to study $P$ from a particular experimental perspective.
For example, one measurement context corresponds to $X=0$ and $Y=1$ which yields probabilities over the four possible outcomes of joint measurements of $A$ and $B$. 
We will denote these outcomes as 
$\{00|01,01|01,10|01,11|01\}$.
For example, $00|01$ denotes the outcome $A=0,B=0$ in the measurement context $M_2=\{X=0,Y=1\}$.

Measurement contexts are formally defined as hyperedges in a hypergraph called a ``contextuality scenario''.
Contextuality scenarios $\mathcal{X}_i, 1\leq i \leq 2$ are composed into a composite contextuality scenario $\mathcal{X}$, which is a hypergraph describing the phenomenon $P$. 
Composition offers the distinct advantage of allowing experimental designs to be theoretically underpinned by hypergraphs in a modular way \cite{bruza:2018}.
More formally, a \emph{contextuality scenario} is a hypergraph $X=(V,E)$ such that:
\vspace*{-0.2cm}
\begin{itemize}
	\item $v \in V$ denotes an outcome which can occur in a measurement context
	\vspace*{-0.2cm}
	\item $e \in E$ is the set of all possible outcomes given a particular measurement context 
\end{itemize}
See Definition 2.2.1 in Ref.~\cite{acin:2015}.

It is important to note that the PPL functions as both a means to simulate an EPR experiment as well as determine whether quantum correlations are present.
As we will see below, each hyperedge of $\mathcal{X}$ is a probability distribution over outcomes in a given measurement context.
In EPR experiments these 
distributions are computed by a sampling process which ensures that the no-signaling constraint is adhered to.   
In order to achieve this, the hypergraphs $\mathcal{X}_i$ are composed using the Foulis--Randall (FR) product \cite{acin:2015} (see the next section).
As a consequence, the PPL must implement this product for a valid simulation of an EPR experiment.
Much of the technical detail to follow describes how this can be achieved. 
To our knowledge the FR product has never been implemented before in a PPL. 
Several such implementations will be specified below in various PPLs and then compared.

At the conclusion of the simulation, the CHSH equalities can be applied to correlations computed from relevant hyperedges in the composite contextuality scenario $\mathcal{X}$ to determine whether quantum correlations are present.
If so, the PPL has successfully simulated phenomenon $P$ as exhibiting quantum, rather than, classical statistics.

\subsection{Foulis--Randall product}

The FR product is used to compose contextuality scenarios as its product ensures no signalling between systems represented by the variables $A$ and $B$ \cite{acin:2015}.

The \emph{Foulis--Randall product} is the scenario $ H_{A} \otimes H_{B} $ with 
\begin{align*}
V\left ( H_{A} \otimes H_{B} \right ) = V(H_{A}) \times V(H_{B}), \; \; \;
E(H_{A} \otimes H_{B}) = E_{A \rightarrow B}  \cup E_{B \rightarrow A}
\end{align*}
where
\begin{align*}
E_{A\rightarrow B}:= \left \{ \bigcup_{a\in e_{A}}  \left \{ a \right \} \times f\left ( a \right ) : e_{A} \in E_{A} , f : e_{A} \rightarrow E_{B} \right  \} \\
\linebreak
E_{A\leftarrow B}:= \left \{ \bigcup_{b\in e_{B}}  \left \{ b \right \} \times f\left ( b \right ) : e_{B} \in E_{B} , f : e_{B} \rightarrow E_{A} \right  \}
\end{align*}
The preceding definition formalizes the simultaneous measurements of the two systems $A$ and $B$ such that no-signalling occurs between these systems \cite{acin:2015}. 
The no-signalling constraint is imposed by means of a set of specific hyperedges which are a consequence of the FR product.

\begin{figure}[h]
\captionsetup{justification=centering}
\centering
\includegraphics[width=8cm]{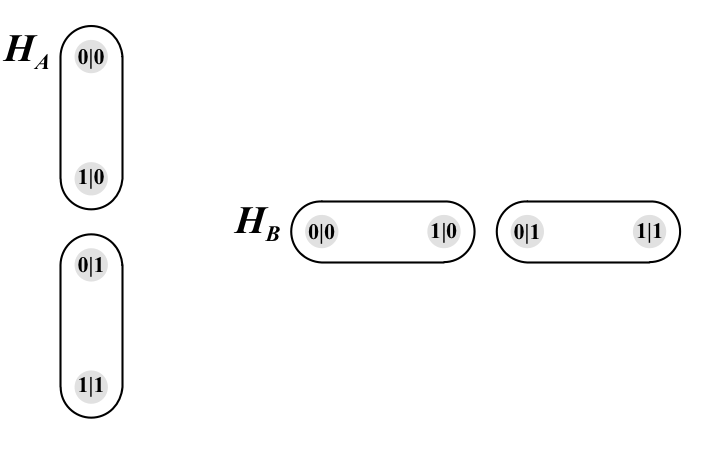}
\caption{Hypergraph Representation Of EPR Systems A \& B}\label{fig:h_a_h_b}
\end{figure}

We now turn to the issue of modularity which was mentioned previously. 
There are two systems $A$ and $B$.
System $A$ has two measurement contexts: 1) $A|X=0$ and 2) $A|X=1$, where both measurements yield an outcome $A=0$ or $A=1$.
In the hypergraph formalism, a measurement context is formalized by a hyperdge.
The hypergraph $H_A$ therefore has two hyperedges, one for each measurement contexts. 
These two hyperedges are visually represented on the LHS of Fig.~\ref{fig:h_a_h_b}. 
Similarly, hypergraph $H_B$ comprises two edges.
$H_A$ and $H_B$ can be viewed as modules which can be composed in various ways to suit the requirements of a particular experimental design. 
In the EPR experiment, four measurement contexts are required in which $A$ are $B$ are jointly measured subject to the no-signalling condition.

\begin{figure}[h]
\captionsetup{justification=centering}
\centering
\includegraphics[width=9cm]{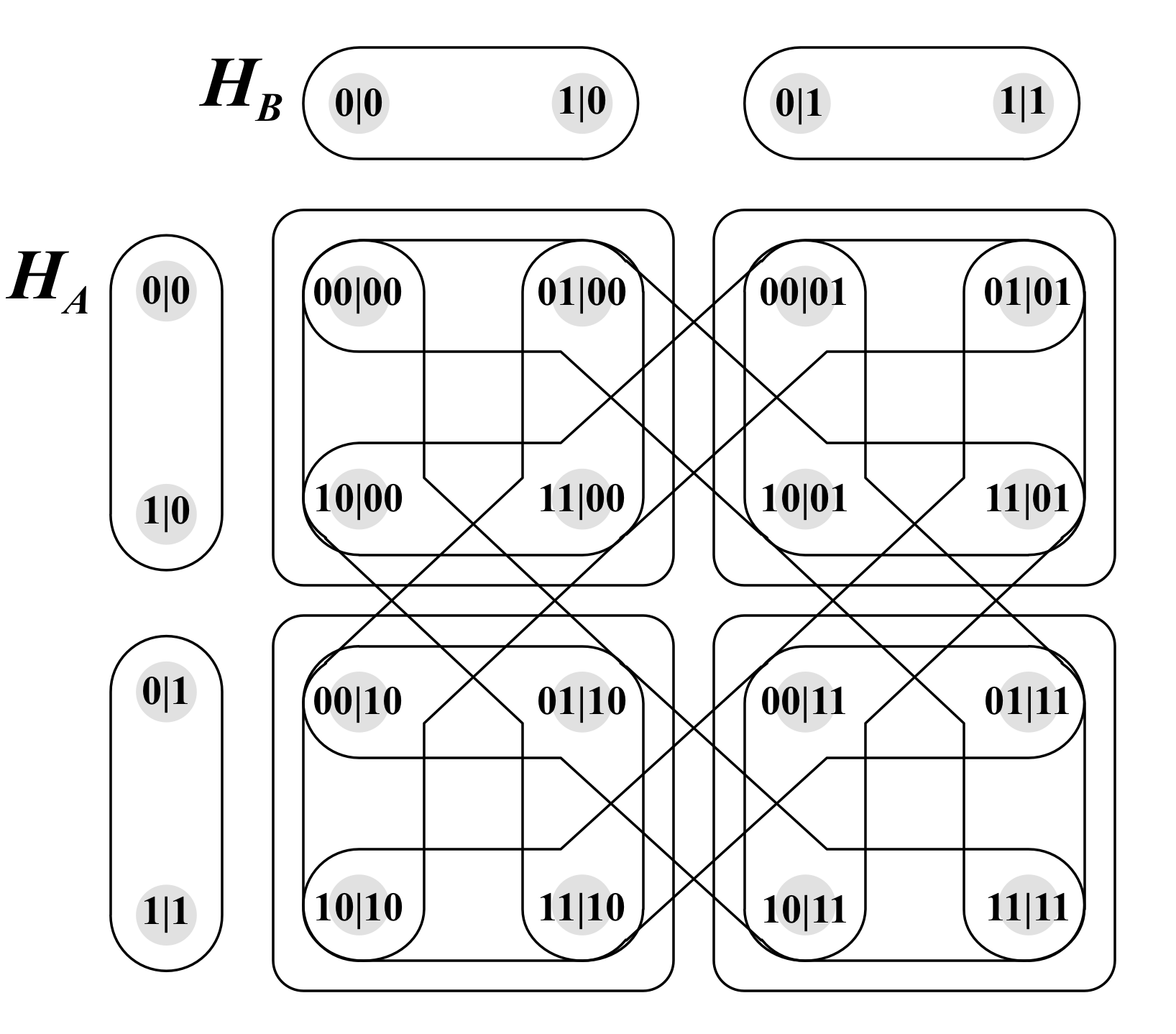}
\caption{Hyperedges Of Foulis--Randall Product}\label{fig:twelve_constraints}
\end{figure}

In order to achieve this, the hypergraphs $H_A$ and $H_B$ are composed using the FR product to produce a composite hypergraph.
The corresponding hypergraph contains 12 edges. Four of these edges correspond to the four pairwise distributions depicted in Fig.~\ref{fig:p16} and 8 additional edges which ensure that no-signalling can occur.
The FR product produces the hypergraph depicted in  Fig.~\ref{fig:twelve_constraints}.
This hypergraph corresponds to composite contextuality scenario $\mathcal{X}$ depicted in Fig.~\ref{fig:PPL-framework}.

To assist with understanding this formalism, one single hyperedge's calculation is considered.

Let $e_{A}$ be equivalent to edge $\left \{  0|0 , 1|0  \right \}$ of hypergraph $H_{A}$

The relevant calculation associated with the instance may then be one of two combinations:

$f\left ( 0|0 \right ) \cup f\left ( 1|0 \right )$ or 
$f\left ( 1|0 \right ) \cup f\left ( 0|0 \right )$ 

The first of the two combinations is selected, expanding to the following expression:

$\left \{  00|00 , 01|00  \right \} \cup \left \{  10|01 , 11|01  \right \}$

\begin{figure}[h]
\captionsetup{justification=centering}
\centering
\includegraphics[width=9cm]{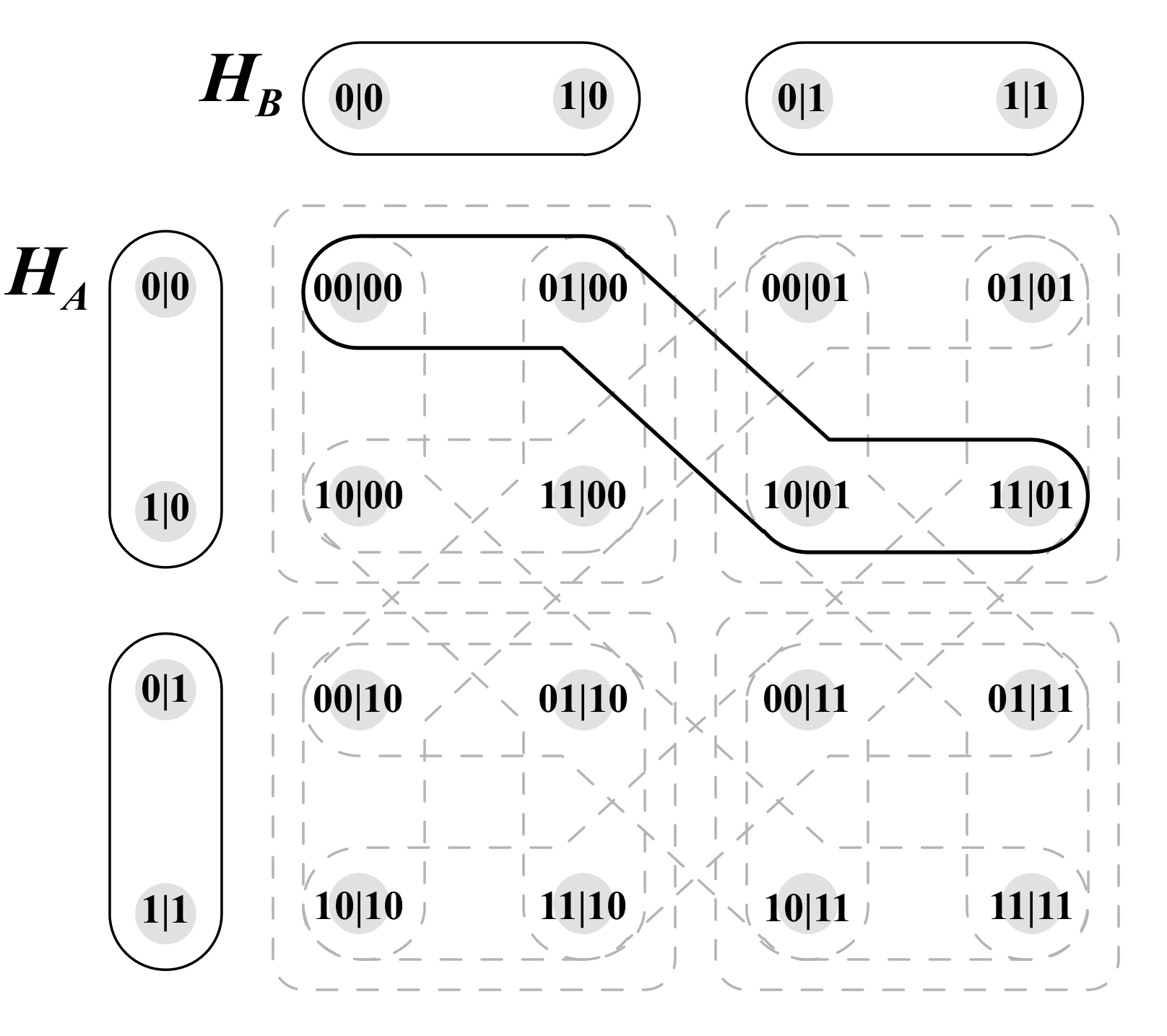}
\caption{Single Hyperedge Of Foulis--Randall Product}\label{fig:foulis_randall_one_hyperedge}
\end{figure}

The hyperedge is isolated in Fig.~\ref{fig:foulis_randall_one_hyperedge}.

In what follows, we implement this hypergraph formalism in several probabilistic programming languages and evaluate the advantages of each.

\section{Implementations}
In this section, four commonly available PPLs illustrate a simulation of the same EPR experiment. 
The goal of this comparison is to judge their relative effectiveness for this purpose.

\subsection{Scope Of Investigation}

Four PPLs were chosen for both qualitative and quantitaive comparison and are listed in Table~\ref{tab:characteristics}.
While other PPLs such as Stan~\cite{carpenter2017stan}, Church~\cite{goodman2008church}, or WebPPL~\cite{goodman:stuhlmuller:2014} were considered for the investigation, we decided to exclude such domain-specific languages on the basis of limited applications in quantum physics. 
Probabilistic programming frameworks that only focus on directed graphs, such as Edward~\cite{tran2016edward}, were also excluded, since this feature is not relevant to the EPR experiment in the hypergraph formalism.

\begin{table}
	\begin{tabular}{llll}
		\textbf{Name} & \textbf{Programming language} & \textbf{License} & \textbf{Supported OS}\\
		\hline
		PyMC3~\cite{salvatier2016probabilistic} & Python & Apache-2.0 & Windows, Mac, Linux \\
		Figaro~\cite{figaro} &  Scala & Custom & Windows, Mac, Linux \\
		Turing.jl~\cite{Turing2016} & Julia & MIT & Windows, Mac, Linux \\
		Pyro~\cite{bingham2018pyro}  & Python & Custom & Windows, Mac, Linux \\
		\hline
	\end{tabular}
\caption{Basic characteristics of probabilistic programming languages}
\label{tab:characteristics}
\end{table}

\subsection{Qualitative Comparison Of PPLs}
The qualitative comparison highlights important pragmatic aspects of probabilistic programs, and is defined by the following criteria.
\subsubsection{Criteria Of Comparison}
\begin{itemize}
  \item \textbf{Extensibility: }The PPL accommodates for simulation of complex experimental settings. This may be inherent in the PPL's means of extension i.e., is open-source, or whether its syntactic constructs provide flexibility in specifying data structures and the flow of control. 
  \item \textbf{Accessibility: }The PPL is intuitive and coherent. Possibly by means of expressive constructs, or comprehensive supporting documentation, accessibility may also be demonstrated by the PPL's community base, or degree of application.
  \item \textbf{Acceleration: }The PPL implements methods of optimization for its execution, reflected in the speed and resource-utilization of its compilation. Acceleration may also be demonstrated in the PPL's scalability. 
\end{itemize}
These criteria are derived from criteria commonly used to judge programming languages.
\subsubsection{Extensibility Of PPLs}

Regarding extensibility, all PPLs are supplemented with repositories containing source code that can be (with respect to licenses) modified and re-compiled. While all PPLs offer containment systems for configuration of probabilistic models, only PyMC3 and Figaro provide tools for the diagnosis and validation of models. In considering Turing.jl's dependence on the Distributions.jl~\cite{distributions_package} package, it can be said that all PPLs provide a number of distribution configurations. In contrast, not all PPLs offer flexibility in step methods used for sampling. This can be overlooked considering that beyond common sampling methods, excess of configurations are typically specialised. Figaro and Pyro are the only PPLs to offer control-flow independence in probabistic inference; both Figaro's inference algorithms and Pyro's strong integration with Python allow for atomic inference processing. Both are also the only PPLs to offer comprehensive open universe simulation\cite{milch2010extending}. All PPLs except Turing.jl provide constructs for the manipulation of the underlying inference algorithms. Table~\ref{tab:extensibility} demonstrates the articulated features in comparison.

\begin{table}
	\begin{tabular}{lllll}
		\textbf{Criteria} & \textbf{PyMC3} & \textbf{Figaro} & \textbf{Turing.jl} & \textbf{Pyro}\\
		\hline
		Control-Flow Independence & \ding{55} & \ding{51} & \ding{55} & \ding{51} \\
        Open Universe Simulation & \ding{55} & \ding{51} & \ding{55} & \ding{51}  \\
        Distribution Configurations & \textbf{$\sim$60} & \textbf{$\sim$36} & \textbf{?} & \textbf{$\sim$39} \\
        Step Methods & \textbf{$\sim$8} & \textbf{$\sim$41} & \textbf{$\sim$13} & \textbf{$\sim$4} \\
        Algorithm Manipulation & \ding{51} & \ding{51} & \ding{55} & \ding{51} \\
        Online Repository & \ding{51} & \ding{51} & \ding{51} & \ding{51} \\
        Model Configuration & \ding{51} & \ding{51} & \ding{51} & \ding{51} \\
        Model Validation & \ding{51} & \ding{51} & \ding{55} & \ding{55} \\
		\hline
	\end{tabular}
\caption{Comparison Of Extensibility Of PPLs}
\label{tab:extensibility}
\end{table}

\subsubsection{Accessibility Of PPLs}

While PyMC3 and Turing.jl have seen a wealth of research projects conducted since their conception, the later debut of Figaro and Pyro has perpetuated fewer examples of application. In light of this, both PPLs provide tutorial literature, and have more comprehensive API reference documentation than the former two. In contrast, Turing.jl has limited tutorial content to support its usage, and does not provide a complete API reference. For ease of use, Pyro advertises its design for agile development, however its syntactic conventions do not warrant any significant differences compared to PyMC3. Nevertheless, both are more easily applied than Figaro or Turing.jl.

\subsubsection{Acceleration Of PPLs}

Concerning acceleration, PyMC3 bases its optimization on Theano's architecture, which is an open-sourced project originally produced at the Université de Montréal \cite{alrfou2016theano}. Correctly applying the Theano architecture with respect to the  GPU on which the PPL is running is a multi-staged process. As Theano depends on the NVIDIA CUDA Developer Toolkit, the GPU's compatibility with the toolkit's contained drivers must be verified before installation can occur. Thereafter, the software `self-validates', and PyMC3 configuration settings must be altered to recognize the GPU support. Only in the instance that the toolkit is correctly installed can PyMC3 take full advantage of its GPU acceleration capabilities. For contrast, no other PPLs evaluated require manual extension of acceleration. For current experimentation, Theano may be suitable, however its discontinuation as of 2017 \cite{peng_2017} poses a threat to using it as a stable basis for future development. For comparison, Figaro was designed specifically for usage within demanding experimental designs. The development team has stressed the library’s capability with its various capabilities e.g. open universe models, spatio-temporal models, recursive models, or infinite models \cite{milch2010extending}.

Similarly, Uber AI Labs stresses that Pyro can be easily scaled to projects of demanding size \cite{pyro_ui_nov2017}; it should be noted that Pyro is based on Pytorch framework, and as a result takes advantage of Pytorch's strong GPU accelerated machine learning tools~\cite{paszke2017pytorch}.

\subsection{Illustration of the EPR Experiment in the four PPLs}

In specifying the EPR experiment in different PPLs varying syntactic constructs can be are highlighted and contrasted, as well as the differing approaches to the simulation.

\subsubsection{PyMC3}

A PyMC3 model defines a context management system used to isolate operations undertaken on stochastic random variables, and thus Python's \verb|with| keyword is applied to automate the release of resources after implementation. Inside the model, the \verb|Bernoulli| method specifies that a distribution of Bernoulli values will be simulated for a given random variable. A probability is also given to direct the sampler towards a bias when generating the distribution. To assist with randomizing results, a \verb|Uniform| distribution is also declared. Then the \verb|sample| method invokes a number of iterations over the specified model. PYMC3 designates tuning of results prior to sampling, as well as indication of a sampling method for which a number of algorithms are offered. In the example, \verb|Metropolis| implies the Metropolis-Hastings algorithm will be used to obtain random results. Upon execution, the model generates a trace containing distributions reflecting the earlier declared random variables.

As this is the first example of code for the experimentation, annotations expressing the meaning of the code are included throughout the implementation.

\begin{lstlisting}[language=python]

 from numpy import zeros, array, fliplr, sum
 from itertools import product
 import pymc3 as pm
%*\color{lgray}{.}*)
\end{lstlisting}

The first block of the implementation declares assistant methods used for value conversions. The first, being the \verb|get_vertex| method linearises the binary variables $X$, $Y$, $A$, and $B$ used to express probabilities of the global distribution into an index of an array.
\begin{align*}
get\_vertex\left ( x ,\ y ,\ a ,\ b \right ) := ((x\times 8)+(y\times 4))+(b+(a\times 2))
\end{align*}

\begin{lstlisting}[language=python]

 def get_vertex(a, b, x, y):
    return ((x*8)+(y*4))+(b+(a*2))
%*\color{lgray}{.}*)
\end{lstlisting}

The second method, \verb|get_hyperedges| leverages enumeration techniques to retrieve hyperedges for contained vertices. 

\begin{lstlisting}[language=python]

 def get_hyperedges(H, n):
    l = []
    for idx, e in enumerate(H):
       if n in e:
          l.append(idx)
    return l
%*\color{lgray}{.}*)
\end{lstlisting}

The \verb|foulis_randall_product| method generates the binary coordinates for all hyperedges in the FR product.

\begin{lstlisting}[language=python]

 def foulis_randall_product():
    fr_edges = []
%*\color{lgray}{.}*)
\end{lstlisting}

The first step involves declaring the hypergraphs for both EPR systems.
\begin{align*}
((((      0 ,\ 0     ) ,\   (     1 ,\ 0   )   )   ,\   ((  0   ,\  1        )   ,\   (     1    ,\  1      )))     ,\    (((     0 ,\  0      )  ,\     (    1 ,\ 0       ))   ,\     ((    0 ,\ 1       )   ,\   (      1  ,\ 1     ))   )) 
\end{align*}

\begin{lstlisting}[language=python]

    H = [   [[[0, 0], [1, 0]], [[0, 1], [1, 1]]], 
            [[[0, 0], [1, 0]], [[0, 1], [1, 1]]]]
%*\color{lgray}{.}*)
\end{lstlisting}

The next step involves producing four hyperedges to represent the four explicit joint measurement contexts on both systems. Two variables are given to assist with computing this result.
\begin{align*}
g \in E_{A} ,\  h \in E_{B}
\end{align*}

\begin{lstlisting}[language=python]

    for edge_a in H[0]:
       for edge_b in H[1]:
%*\color{lgray}{.}*)
\end{lstlisting}

Thereafter, each hyperedge is defined as the combined sets produced by the following expression.
\begin{align*}
(       \forall i \in g  ,\      \forall j \in h :   \forall w \in i  ,\      \forall x \in j :  (   w_{1} ,\   x_{1} ,\ w_{2} ,\ x_{2} ) )     
\end{align*}

\begin{lstlisting}[language=python]

          fr_edge = []
          for vertex_a in edge_a:
             for vertex_b in edge_b:
                fr_edge.append([   
                    vertex_a[0], vertex_b[0], 
                    vertex_a[1], vertex_b[1]])
          fr_edges.append(fr_edge)
%*\color{lgray}{.}*)
\end{lstlisting}

The last step involves calculating the hyperedges of both systems as dependent on the other. To achieve this programmatically, three variables are declared. The first ($m$) are all members of set $M$, where $M$ are the possible measurement choices for the scenario (in this case two), the second ($n$) being the other possible measurement choice, and the last variable $o$ being all edges from the hypergraph associated with the measurement choice.

\begin{align*}
\forall m \in M : n = m' ,\ \forall o \in E(H_{m})
\end{align*}

\begin{lstlisting}[language=python]

    for mc in range(0,2):
       mc_i = abs(1-mc)
       for edge in H[mc]:
%*\color{lgray}{.}*)
\end{lstlisting}

For each $o$, in some selected hypergraph, a second variable $j$ is declared as two possible values. For each possibility, a hyperedge is then defined as the variable $k$, being the combination of all vertices in $o$ given to a function that produces the hyperedge.

\begin{align*}
\forall j \in ( 1 ,\ 2 ) : k = ( \forall l \in E(H_{n}) : f \left ( l ,\ j ,\ m ,\ n ,\ o \right )  )
\end{align*}

\begin{lstlisting}[language=python]

          for j in range(0,2):
             fr_edge = []
             for i in range(0, len(edge)):
%*\color{lgray}{.}*)
\end{lstlisting}

The mentioned function computes the hyperedge by declaring single-use variables $s$, $q$, $r$, $u$, $t$, and $v$.

\begin{align*}
E(H_{n})_s = l' ,\ q = o_{|s-j|+1} ,\ r = l_{1} ,\ u = l_{2} ,\ t = \left ( q_{1}  ,\ r_{1} ,\ q_{2}  ,\ r_{2} \right ) ,\ v = \left ( q_{1}  ,\ u_{1} ,\ q_{2}  ,\ u_{2} \right ) 
\end{align*}

\begin{lstlisting}[language=python]

                edge_b = H[mc_i][i]
                vertex_a = edge[abs(i-j)]
                vertex_b = edge_b[0]
                vertex_c = edge_b[1]
                vertices_a = [
                   vertex_a[0], vertex_b[0], 
                   vertex_a[1], vertex_b[1]]
                vertices_b = [
                   vertex_a[0], vertex_c[0], 
                   vertex_a[1], vertex_c[1]]
%*\color{lgray}{.}*)
\end{lstlisting}

Thereafter, a set is constructed by use of its variables, and portions of the desired hyperedges are iteratively returned.

\begin{align*}
\left ( \left ( t_{m} ,\ t_{n} ,\ t_{m+2} ,\ t_{n+2} \right ) ,\ \left ( v_{m} ,\ v_{n} ,\ v_{m+2} ,\ v_{n+2} \right ) \right )
\end{align*}

Upon calculation of the last step of the process, the hyperedges corresponding to the measurement choices of both EPR systems as dependent on the other are produced, totaling the necessary constraints described in binary format.

\begin{lstlisting}[language=python]

                fr_edge.append([
                   vertices_a[mc], vertices_a[mc_i], 
                   vertices_a[mc+2], vertices_a[mc_i+2]]
                   )
                fr_edge.append([
                   vertices_b[mc], vertices_b[mc_i], 
                   vertices_b[mc+2], vertices_b[mc_i+2]])
             fr_edges.append(fr_edge)
    return fr_edges
%*\color{lgray}{.}*)
\end{lstlisting}

\begin{figure}[h]
	\captionsetup{justification=centering}
	\centering
	\includegraphics[width=9cm]{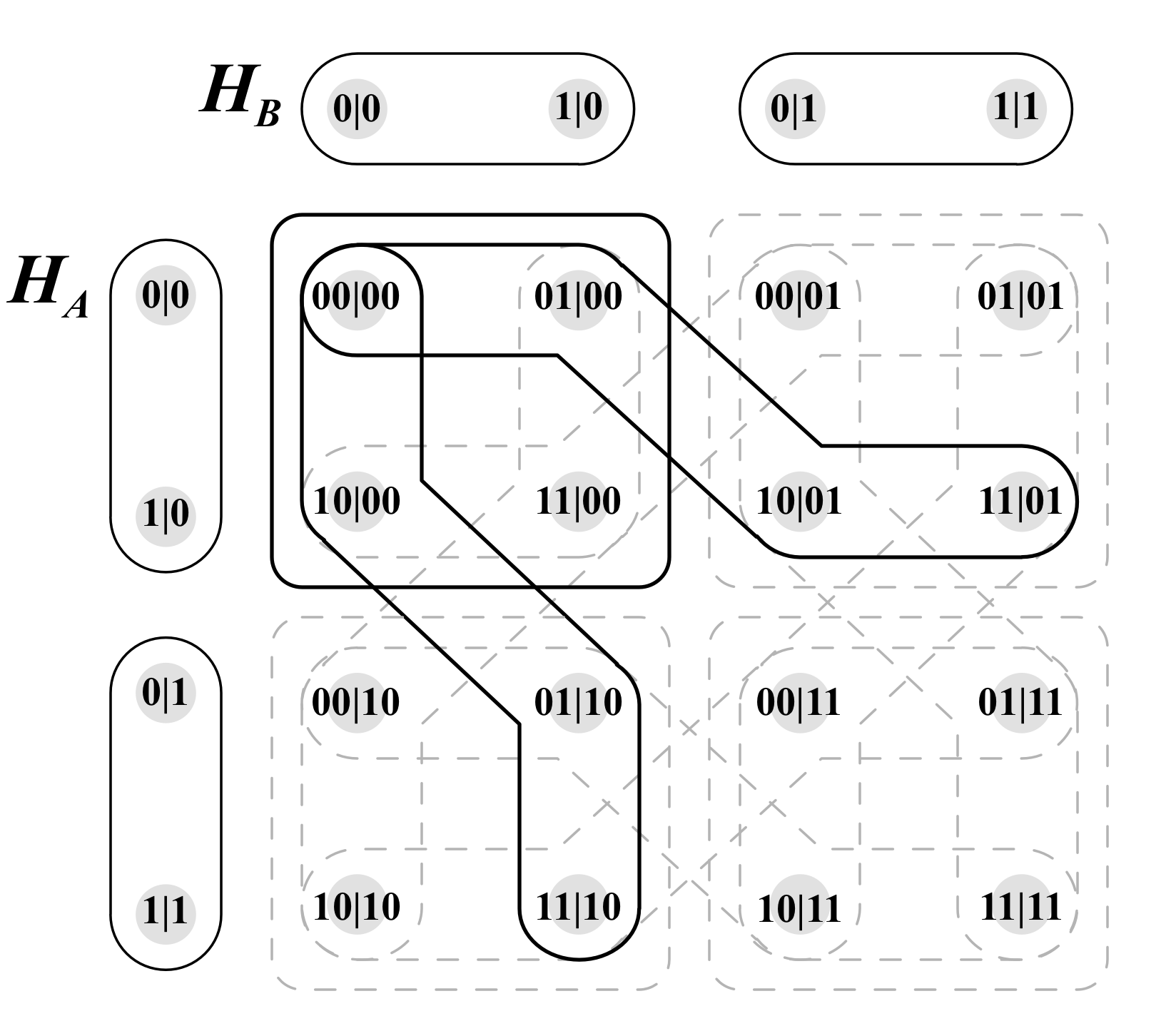}
	\caption{Visualisation Of Hyperedges Associated With Vertex 00$|$00}\label{fig:foulis_randall_normalised}
\end{figure}

To compute the four pairwise distributions at the basis of the EPR experiment (See Fig.~\ref{fig:p16}, an iterative sampling process is undertaken for the four variables $a$, $b$, $x$, and $y$ that were previously mentioned as specifying one of the 16 possible vertices. 
These values are restricted to binary outcomes by means of specifying Bernoulli distributions for which the sampler runs the process. 
The experiment is fixed such that each vertex has an equal possibility of being sampled as any other vertex, and results may only be discounted if they do not comply with specified input correlations.
Upon selecting a vertex at a step in the iterative process, the array index associated with the binary representation is incremented by one, via the use of the vertex mapping function. Simultaneously, another array representing the hyperedges of the FR product is also incremented by one at all indexes associated with the hyperedges containing the said vertex. The iterative process only exits when the sum of the global distribution is equivalent to the desired number of iterations. Thereafter, each vertex is normalised by the sum of the values in the corresponding array of hyperedges in which its associated vertex is contained, and is multiplied by 3 (for reflection of the number of associated hyperedges). A visualisation of the hyperedges associated with the vertex at index $00|00$ of the global distribution can be seen in Fig.~\ref{fig:foulis_randall_normalised}.

If the said vertex sustains a weight of 10, and the combined weight of its associated hyperedges is 40, the normalised weight of the vertex will equate to 0.75.

\begin{lstlisting}[language=python]

 def generate_global_distribution(constraints,N):
    hyperedges = foulis_randall_product()
    hyperedges_tallies = zeros(12)
    global_distribution = zeros(16)
    while sum(global_distribution) < N:
       with pm.Model():
          pm.Uniform('C',0.0,1.0)
          pm.Bernoulli('A',0.5)
          pm.Bernoulli('B',0.5)
          pm.Bernoulli('X',0.5)
          pm.Bernoulli('Y',0.5)
          S = pm.sample(N,tune=0, step=pm.Metropolis())
          c = S.get_values('C')
          a = S.get_values('A')
          b = S.get_values('B')
          x = S.get_values('X')
          y = S.get_values('Y')
       for i in range(0, N):
          if (c[i] < constraints[x[i]][y[i]][a[i],b[i]]):
             for edge in get_hyperedges(hyperedges, 
                [a[i], b[i], x[i], y[i]]):
                hyperedges_tallies[edge] += 1
             global_distribution[
                get_vertex(a[i], b[i], x[i], y[i])] += 1
    z = [0,1]
    for a, b, x, y in product(z,z,z,z):
       summed_tally = (sum(hyperedges_tallies[e] 
          for e in get_hyperedges(hyperedges, [a, b, x, y])))
       global_distribution[get_vertex(a, b, x, y)] /= summed_tally
    global_distribution *= 3
    return global_distribution
%*\color{lgray}{.}*)
\end{lstlisting}

Given below in Listing~\ref{fig:pymc3_implementation} is the complete undivided implementation of the EPR experimentation in PyMC3.

\begin{lstlisting}[language=python,caption={PyMC3 Implementation Of EPR Simulation},captionpos=b,label={fig:pymc3_implementation}]

 from numpy import zeros, array, fliplr, sum
 from itertools import product
 import pymc3 as pm
 
 def get_vertex(a, b, x, y):
    return ((x*8)+(y*4))+(b+(a*2))

 def get_hyperedges(H, n):
    l = []
    for idx, e in enumerate(H):
       if n in e:
          l.append(idx)
    return l
    
 def foulis_randall_product():
    fr_edges = []
    H = [   [[[0, 0], [1, 0]], [[0, 1], [1, 1]]], 
            [[[0, 0], [1, 0]], [[0, 1], [1, 1]]]]
    for edge_a in H[0]:
       for edge_b in H[1]:
          fr_edge = []
          for vertex_a in edge_a:
             for vertex_b in edge_b:
                fr_edge.append([   
                    vertex_a[0], vertex_b[0], 
                    vertex_a[1], vertex_b[1]])
          fr_edges.append(fr_edge)
    for mc in range(0,2):
       mc_i = abs(1-mc)
       for edge in H[mc]:
          for j in range(0,2):
             fr_edge = []
             for i in range(0, len(edge)):
                edge_b = H[mc_i][i]
                vertex_a = edge[abs(i-j)]
                vertex_b = edge_b[0]
                vertex_c = edge_b[1]
                vertices_a = [
                   vertex_a[0], vertex_b[0], 
                   vertex_a[1], vertex_b[1]]
                vertices_b = [
                   vertex_a[0], vertex_c[0], 
                   vertex_a[1], vertex_c[1]]
                fr_edge.append([
                   vertices_a[mc], vertices_a[mc_i], 
                   vertices_a[mc+2], vertices_a[mc_i+2]]
                   )
                fr_edge.append([
                   vertices_b[mc], vertices_b[mc_i], 
                   vertices_b[mc+2], vertices_b[mc_i+2]])
             fr_edges.append(fr_edge)
    return fr_edges

 def generate_global_distribution(constraints,N):
    hyperedges = foulis_randall_product()
    hyperedges_tallies = zeros(12)
    global_distribution = zeros(16)
    while sum(global_distribution) < N:
       with pm.Model():
          pm.Uniform('C',0.0,1.0)
          pm.Bernoulli('A',0.5)
          pm.Bernoulli('B',0.5)
          pm.Bernoulli('X',0.5)
          pm.Bernoulli('Y',0.5)
          S = pm.sample(N,tune=0, step=pm.Metropolis())
          c = S.get_values('C')
          a = S.get_values('A')
          b = S.get_values('B')
          x = S.get_values('X')
          y = S.get_values('Y')
       for i in range(0, N):
          if (c[i] < constraints[x[i]][y[i]][a[i],b[i]]):
             for edge in get_hyperedges(hyperedges, 
                [a[i], b[i], x[i], y[i]]):
                hyperedges_tallies[edge] += 1
             global_distribution[
                get_vertex(a[i], b[i], x[i], y[i])] += 1
    z = [0,1]
    for a, b, x, y in product(z,z,z,z):
       summed_tally = (sum(hyperedges_tallies[e] 
          for e in get_hyperedges(hyperedges, [a, b, x, y])))
       global_distribution[get_vertex(a, b, x, y)] /= summed_tally
    global_distribution *= 3
    return global_distribution
%*\color{lgray}{.}*)
\end{lstlisting}

\subsubsection{Turing.jl}

Like PyMC3, Turing.jl isolates operations on random variables to a single model with the use of the \verb|@model| macro. To obtain randomly sampled non-negative values for a Bernoulli distribution, the model requires the declaration of a uniform Beta prior, invoked with the \verb|Beta| method. Then Bernoulli distributions are declared with a \verb|Bernoulli| method, once again accompanied by probabilities describing sampling biases for later generated distributions, as well as a \verb|Uniform| distribution.

In the \verb|generate_global_distribution| method of Listing~\ref{fig:turing_implementation}, the \verb|sample| function invokes the model, a step-method, as well as the number of desired iterations. In this case, the Sequential Monte Carlo sampling (abbreviated to \verb|SMC|) has been applied.

To obtain the trace of a distribution in the model, the output must be indexed with the precession of a colon. In the example, the results are retrieved, tallied, and normalised by means of the \verb|foulis_randall_product| method, before returning the result.

\begin{lstlisting}[language=java,caption={Turing.jl Implementation Of EPR Simulation},captionpos=b,label={fig:turing_implementation}]

 %*\color{red}{using}*) Turing
 %*\color{red}{using}*) Distributions
 
 %*\color{red}{function}*) %*\color{darkolivegreen}{foulis\_randall\_product}*)()
    fr_edges %*\color{red}{=}*) Array{Array{Array{Float64}}}(0)
    H %*\color{red}{=}*) [    [[[0.0,0.0],[1.0,0.0]],[[0.0,1.0],[1.0,1.0]]],
             [[[0.0,0.0],[1.0,0.0]],[[0.0,1.0],[1.0,1.0]]]]
    %*\color{red}{for}*) i %*\color{red}{=}*) 1%*\color{red}{:}*)%*\color{darkolivegreen}{size}*)(H[1])[1]
       %*\color{red}{for}*) j %*\color{red}{=}*) 1%*\color{red}{:}*)%*\color{darkolivegreen}{size}*)(H[2])[1]
          fr_edge %*\color{red}{=}*) Array{Array{Float64}}(0)
          %*\color{red}{for}*) k %*\color{red}{=}*) 1%*\color{red}{:}*)%*\color{darkolivegreen}{size}*)(H[    1][i])[1]
             %*\color{red}{for}*) l %*\color{red}{=}*) 1%*\color{red}{:}*)%*\color{darkolivegreen}{size}*    )(H[1][j])[1]
                append!( fr_edge, 
                [[  H[1][i][k][1],H[2][j][l][1] , 
                    H[1][i][k][2],H[2][j][l][2] ]] )
             %*\color{red}{end}*)
          %*\color{red}{end}*)
          append!(fr_edges,[fr_edge])
       %*\color{red}{end}*)
    %*\color{red}{end}*)
    %*\color{red}{for}*) mc %*\color{red}{=}*) 1%*\color{red}{:}*)2
       mc_i %*\color{red}{=}*) %*\color{darkolivegreen}{abs}*)(3%*\color{red}{-}*)mc)
       %*\color{red}{for}*) k %*\color{red}{=}*) 1%*\color{red}{:}*)%*\color{darkolivegreen}{size}*)(H[mc])[1]
          %*\color{red}{for}*) j %*\color{red}{=}*) 1%*\color{red}{:}*)2
             fr_edge = Array{Array{Float64}}(0)
             %*\color{red}{for}*) i %*\color{red}{=}*) 1%*\color{red}{:}*)%*\color{darkolivegreen}{size}*    )(H[mc][k])[1]
                edge_b %*\color{red}{=}*) H[mc_i][i]
                vertex_a %*\color{red}{=}*) H[mc][k][%*\color{darkolivegreen}{abs}*)(i%*\color{red}{-}*)j)%*\color{red}{+}*)1]
                vertex_b %*\color{red}{=}*) edge_b[1]
                vertex_c %*\color{red}{=}*) edge_b[2]
                vertices_a %*\color{red}{=}*) [  vertex_a[1], vertex_b[1], 
                   vertex_a[2], vertex_b[2]]
                vertices_b %*\color{red}{=}*) [  vertex_a[1], vertex_c[1], 
                   vertex_a[2], vertex_c[2]]
                this_edge_b %*\color{red}{=}*) Array{Float64}(0)
                append!(fr_edge,[[ 
                   vertices_a[mc],   vertices_a[mc_i], 
                   vertices_a[mc%*\color{red}{+}*)2], vertices_a[mc_i%*\color{red}{+}*)2] ]] )
                append!(fr_edge,[[ 
                   vertices_b[mc],   vertices_b[mc_i], 
                   vertices_b[mc%*\color{red}{+}*)2], vertices_b[mc_i%*\color{red}{+}*)2] ]] )
             %*\color{red}{end}*)
             append!(fr_edges,[fr_edge])
          %*\color{red}{end}*)
       %*\color{red}{end}*)
    %*\color{red}{end}*)
    fr_edges
 %*\color{red}{end}*)
 
 %*\color{red}{function}*) %*\color{darkolivegreen}{get\_vertex}*)(a,b,x,y)
    ((x%*\color{red}{×}*)8)%*\color{red}{+}*)(y%*\color{red}{×}*)4))%*\color{red}{+}*)(b%*\color{red}{+}*)(a%*\color{red}{×}*)2))%*\color{red}{+}*)1
 %*\color{red}{end}*)
 
 %*\color{red}{function}*) %*\color{darkolivegreen}{get\_hyperedges}*)(H, n)
    l %*\color{red}{=}*) []
    %*\color{red}{for}*) i %*\color{red}{=}*) 1%*\color{red}{:}*)%*\color{darkolivegreen}{size}*)(H)[1]
       if %*\color{darkolivegreen}{any}*)(x%*\color{red}{→}*)x%*\color{red}{==}*)n, H[i])
          append!(l,i)
       %*\color{red}{end}*)
    %*\color{red}{end}*)
    l
 %*\color{red}{end}*)
 
 @model %*\color{darkolivegreen}{mdl}*)() %*\color{red}{=}*) %*\color{red}{begin}*)
    z %*\color{red}{≈}*) %*\color{darkolivegreen}{Beta}*)(1,1)
    a %*\color{red}{≈}*) %*\color{darkolivegreen}{Bernoulli}*)(0.5)
    b %*\color{red}{≈}*) %*\color{darkolivegreen}{Bernoulli}*)(0.5)
    x %*\color{red}{≈}*) %*\color{darkolivegreen}{Bernoulli}*)(0.5)
    y %*\color{red}{≈}*) %*\color{darkolivegreen}{Bernoulli}*)(0.5)
    c %*\color{red}{≈}*) %*\color{darkolivegreen}{Uniform}*)(0.0, 1.0)
 %*\color{red}{end}*)
 
 %*\color{red}{function}*) %*\color{darkolivegreen}{generate\_global\_distribution}*)(constraints,N)
    hyperedges %*\color{red}{=}*) %*\color{darkolivegreen}{foulis\_randall\_product}*)()
    hyperedges_tallies %*\color{red}{=}*) %*\color{darkolivegreen}{zeros}*)(12)
    global_distribution %*\color{red}{=}*) %*\color{darkolivegreen}{zeros}*)(16)
    %*\color{red}{while}*) %*\color{darkolivegreen}{sum}*)(global_distribution) < N
       r %*\color{red}{=}*) %*\color{darkolivegreen}{sample}*)(mdl(), %*\color{darkolivegreen}{SMC}*)(N))
       a %*\color{red}{=}*) r[%*\color{red}{:}*)a]
       b %*\color{red}{=}*) r[%*\color{red}{:}*)b]
       x %*\color{red}{=}*) r[%*\color{red}{:}*)x]
       y %*\color{red}{=}*) r[%*\color{red}{:}*)y]
       c %*\color{red}{=}*) r[%*\color{red}{:}*)c]
       %*\color{red}{for}*) i %*\color{red}{=}*) 1%*\color{red}{:}*)N
          %*\color{red}{if}*) (c[i] < constraints[x[i]%*\color{red}{+}*)1][y[i]%*\color    {red}{+}*)1][a[i]%*\color{red}{+}*)1][b[i]%*\color{red}{+}*)1])
             I %*\color{red}{=}*) [%*\color{darkolivegreen}{convert}*)(Float64,a[i]), %*\color    {darkolivegreen}{convert}*)(Float64,b[i]), 
                  %*\color{darkolivegreen}{convert}*)(Float64,x[i]), %*\color{darkolivegreen}{convert}*)(Float64,y[i])]
             associated_hyperedges %*\color{red}{=}*) %*\color{darkolivegreen}{get\_hyperedges}*)(hyperedges, I)
             %*\color{red}{for}*) j %*\color{red}{=}*) 1%*\color{red}{:}*)%*\color{darkolivegreen}{size}*    )(associated_hyperedges)[1]
                hyperedges_tallies[
                   associated_hyperedges[j]] %*\color{red}{+=}*) 1
             %*\color{red}{end}*)
             global_distribution[
                %*\color{darkolivegreen}{get\_vertex}*)(a[i], b[i], x[i], y[i])] %*\color{red}{+=}*) 1
          %*\color{red}{end}*)
       %*\color{red}{end}*)
    %*\color{red}{end}*)
    %*\color{red}{for}*) a %*\color{red}{=}*) 0%*\color{red}{:}*)1, b %*\color{red}{=}*) 0%*\color{red}{:}*)1, x %*\color{red}{=}*)    0%*\color{red}{:}*)1,  y %*\color{red}{=}*) 0%*\color{red}{:}*)1
       summed_amount %*\color{red}{=}*) 0
       I %*\color{red}{=}*) [ %*\color{darkolivegreen}{convert}*)(Float64,a), %*\color{darkolivegreen}{    convert}*)(Float64,b), 
             %*\color{darkolivegreen}{convert}*)(Float64,x), %*\color{darkolivegreen}{convert}*)(Float64,y) ]
       associated_hyperedges %*\color{red}{=}*) %*\color{darkolivegreen}{get\_hyperedges}*)(hyperedges, I)
       %*\color{red}{for}*) edge_index %*\color{red}{=}*) 1%*\color{red}{:}*)size(associated_hyperedges)[1]
          summed_amount %*\color{red}{+=}*) hyperedges_tallies[edge_index]
       %*\color{red}{end}*)
       global_distribution[%*\color{darkolivegreen}{get\_vertex}*)(a, b, x, y)] %*\color{red}{/=}*) summed_amount
    %*\color{red}{end}*)
    global_distribution %*\color{red}{.×}*) 3
 %*\color{red}{end}*) 
%*\color{lgray}{.}*)
\end{lstlisting}
\subsubsection{Figaro}

To achieve a joint-probability distribution on a measurement context of random variables, Figaro's syntactic elements reveal fundamental differences in its approach. A class is the advised object for the purpose of declaring a model. For each random variable in the model, a probability bias is applied to Figaro's \verb|Flip| method, generating a Bernoulli distribution on which the \verb|If| method can then associate the results to desired values, or perpetuation of other methods.

In Listing~\ref{fig:figaro_implementation}, states are bound to integers. Thereafter, possible joint outcomes of random variables are permuted through articulation of expressions concerning the previously mentioned states. In the \verb|GenerateGlobalDistribution| method, it can be seen that after initialising the FR product (via the \verb|FoulisRandallProduct| method), the sampling process is called by means of \verb|start|, \verb|stop|, and \verb|kill| chains applied on the \verb|algorithm| object. On the preceding line, the \verb|MetropolisHastings| method implies the Metropolis-Hastings step-method will be used for the sampling process, and the \verb|outcomes| of the \verb|model| class will be considered. For more complex experiments, the \verb|ProposalScheme| may be modified, however not in this case. The \verb|sampleFromPosterior| sub-method chained to calls on each variable compile the required distributions on execution. The \verb|take| sub-method chained to the sampling methods are used to declare the number of outcomes retrieved from the sampler. This aspect is consequent of sampler delivering results via \verb|Stream| primitives, a resource-efficient consideration ensuring that only required data is evaluated. Thereafter, the proceeding code tallies the indexes of the \verb|globalDistribution| array, and normalises the results.

\begin{lstlisting}[caption={Figaro Implementation Of EPR Simulation},captionpos=b,label={fig:figaro_implementation}]
 
 %*\color{scala-orange}{import}*) com.cra.figaro.algorithm.sampling._
 %*\color{scala-orange}{import}*) com.cra.figaro.language._
 %*\color{scala-orange}{import}*) com.cra.figaro.library.compound.If
 %*\color{scala-orange}{import}*) com.cra.figaro.library.atomic.continuous.Uniform
 
 %*\color{scala-orange}{object}*) Main {
    
    %*\color{scala-orange}{def}*) %*\color{scala-yellow}{FoulisRandallProduct}*)(): Array[Array[Array[%*\color    {scala-orange}{Double}*)]]] = {
       %*\color{scala-orange}{var}*) foulisRandallEdges = Array[Array[Array[%*\color{scala-orange}{Double}*)]]]()
       %*\color{scala-orange}{val}*) hypergraphs = Array(
          Array(  Array( Array(%*\color{scala-blue}{0.0}*),%*\color{scala-blue}{0.0}*)), Array(%*\color    {scala-blue}{1.0}*),%*\color{scala-blue}{0.0}*))    ),
                  Array( Array(%*\color{scala-blue}{0.0}*),%*\color{scala-blue}{1.0}*)), Array(%*\color    {scala-blue}{1.0}*),%*\color{scala-blue}{1.0}*)) ) ),
          Array(  Array( Array(%*\color{scala-blue}{0.0}*),%*\color{scala-blue}{0.0}*)), Array(%*\color    {scala-blue}{1.0}*),%*\color{scala-blue}{0.0}*))    ),
                  Array( Array(%*\color{scala-blue}{0.0}*),%*\color{scala-blue}{1.0}*)), Array(%*\color    {scala-blue}{1.0}*),%*\color{scala-blue}{1.0}*)) ) ) )
       %*\color{scala-orange}{for}*) (edgeA <- hypergraphs(%*\color{scala-blue}{0}*))) {
          %*\color{scala-orange}{for}*) (edgeB <- hypergraphs(%*\color{scala-blue}{1}*))) {
             %*\color{scala-orange}{var}*) foulisRandallEdge = Array[Array[%*\color{scala-orange}{Double}*)]]()
             %*\color{scala-orange}{for}*) (vertexA <- edgeA) {
                %*\color{scala-orange}{for}*) (vertexB <- edgeB) {
                   foulisRandallEdge ++= Array(Array[%*\color{scala-orange}{Double}*)](
                      vertexA(%*\color{scala-blue}{0}*)), vertexB(%*\color{scala-blue}{0}*)), vertexA(%*\color    {scala-blue}{1}*)), vertexB(%*\color{scala-blue}{1}*)) ))
                }
             }
             foulisRandallEdges ++= Array(foulisRandallEdge)
          }
       }
       %*\color{scala-orange}{for}*) ( measurementChoice <- %*\color{scala-blue}{0}*) to %*\color{scala-blue}{1    }*)){
          %*\color{scala-orange}{val}*) measurementChoiceInverse = %*\color{scala-blue}{1}*) - measurementChoice
          %*\color{scala-orange}{for}*) (edge <- hypergraphs(measurementChoice)) {
             %*\color{scala-orange}{for}*) ( j <- %*\color{scala-blue}{0}*) to %*\color{scala-blue}{1}*)){
                %*\color{scala-orange}{var}*) foulisRandallEdge = Array[Array[%*\color{scala-orange}{Double}*)]]()
                %*\color{scala-orange}{for}*) ( i <- edge.indices) {
                   %*\color{scala-orange}{val}*) edgeB = hypergraphs(measurementChoiceInverse)(i)
                   %*\color{scala-orange}{val}*) vertexA = edge(Math.abs(i-j))
                   %*\color{scala-orange}{val}*) vertexB = edgeB(%*\color{scala-blue}{0}*))
                   %*\color{scala-orange}{val}*) vertexC = edgeB(%*\color{scala-blue}{1}*))
                   %*\color{scala-orange}{val}*) verticesA = Array(  vertexA(%*\color{scala-blue}{0}*)), vertexB(%*\color{scala-blue}{0}*)),
                                          vertexA(%*\color{scala-blue}{1}*)), vertexB(%*\color{scala-blue}{1}*)) )
                   %*\color{scala-orange}{val}*) verticesB = Array(  vertexA(%*\color{scala-blue}{0}*)), vertexC(%*\color{scala-blue}{0}*)),
                                          vertexA(%*\color{scala-blue}{1}*)), vertexC(%*\color{scala-blue}{1}*)) )
                   foulisRandallEdge ++= Array(
                      Array[%*\color{scala-orange}{Double}*)](
                         verticesA(measurementChoice),
                         verticesA(measurementChoiceInverse),
                         verticesA(measurementChoice+%*\color{scala-blue}{2}*)),
                         verticesA(measurementChoiceInverse+%*\color{scala-blue}{2}*)) ) )
                   foulisRandallEdge ++= Array(
                      Array[%*\color{scala-orange}{Double}*)](
                         verticesB(measurementChoice),
                         verticesB(measurementChoiceInverse),
                         verticesB(measurementChoice+%*\color{scala-blue}{2}*)),
                         verticesB(measurementChoiceInverse+%*\color{scala-blue}{2}*)) ) )
                }
                foulisRandallEdges ++= Array(foulisRandallEdge)
             }
          }
       }
       foulisRandallEdges
    }
    
    %*\color{scala-orange}{class}*) Model() {
       %*\color{scala-orange}{var}*) %*\color{scala-purple}{outcomes}*) = Array[Element[%*\color{scala-orange}{    Double}*)]]()
       %*\color{scala-orange}{for}*) (i <- %*\color{scala-blue}{0}*) to %*\color{scala-blue}{3}*)) {
          %*\color{scala-purple}{outcomes}*) :+= If(Flip(%*\color{scala-blue}{0.5}*)), %*\color{scala-blue}{0.0}    *), %*\color{scala-blue}{1.0}*))
       }
       %*\color{scala-purple}{outcomes}*) :+= Uniform(%*\color{scala-blue}{0.0}*), %*\color{scala-blue}{1.0}*))
    }
    
    %*\color{scala-orange}{def}*) %*\color{scala-yellow}{GetVertex}*)(a: %*\color{scala-orange}{Int}*), b: %*\color{scala-orange}{Int}*), x: %*\color{scala-orange}{Int}*), y: %*\color{scala-orange}{Int}*)): %*\color{scala-orange}{Int}*) = {
       ((x*%*\color{scala-blue}{8}*))+(y*%*\color{scala-blue}{4}*)))+(b+(a*%*\color{scala-blue}{2}*)))
    }
    
    %*\color{scala-orange}{def}*) %*\color{scala-yellow}{GetHyperedges}*)(H: Array[Array[Array[%*\color    {scala-orange}{Double}*)]]], 
    n: Array[%*\color{scala-orange}{Double}*)]): Array[%*\color{scala-orange}{Int}*)] = {
       %*\color{scala-orange}{var}*) l = Array[%*\color{scala-orange}{Int}*)]()
       %*\color{scala-orange}{for}*) (i <- H.indices) {
          %*\color{scala-orange}{if}*) (H(i).deep.contains(n.deep)) {
             l :+= i
          }
       }
       l
    }
    
    %*\color{scala-orange}{def}*) %*\color{scala-yellow}{GenerateGlobalDistribution}*)(constraints: 
       Array[Array[Array[Array[%*\color{scala-orange}{Double}*)]]]], N: %*\color{scala-orange}{Int}*)): %*\color{scala-orange}{Unit}*) = {
       %*\color{scala-orange}{val}*) hyperedges = FoulisRandallProduct()
       %*\color{scala-orange}{var}*) hyperedgesTallies = Array[%*\color{scala-orange}{Double}*)].fill(%*\color           {scala-blue}{12}*)){%*\color{scala-blue}{0.0}*)}
       %*\color{scala-orange}{var}*) globalDistribution = Array[%*\color{scala-orange}{Double}*)].fill(%*\color           {scala-blue}{16}*)){%*\color{scala-blue}{0.0}*)}
       
       %*\color{scala-orange}{while}*) (globalDistribution.sum < N) {
          %*\color{scala-orange}{var}*) model = %*\color{scala-orange}{new}*) Model()
          %*\color{scala-orange}{val}*) algorithm = MetropolisHastings(N, 
          ProposalScheme.default, model.%*\color{scala-purple}{outcomes}*): _*)
          algorithm.start()
          algorithm.stop()
          algorithm.kill()
          %*\color{scala-orange}{val}*) a = algorithm.sampleFromPosterior(
             model.%*\color{scala-purple}{outcomes}*)(%*\color{scala-blue}{0}*))).take(N).toArray
          %*\color{scala-orange}{val}*) b = algorithm.sampleFromPosterior(
             model.%*\color{scala-purple}{outcomes}*)(%*\color{scala-blue}{1}*))).take(N).toArray
          %*\color{scala-orange}{val}*) x = algorithm.sampleFromPosterior(
             model.%*\color{scala-purple}{outcomes}*)(%*\color{scala-blue}{2}*))).take(N).toArray
          %*\color{scala-orange}{val}*) y = algorithm.sampleFromPosterior(
             model.%*\color{scala-purple}{outcomes}*)(%*\color{scala-blue}{3}*))).take(N).toArray
          %*\color{scala-orange}{val}*) c = algorithm.sampleFromPosterior(
             model.%*\color{scala-purple}{outcomes}*)(%*\color{scala-blue}{4}*))).take(N).toArray
          %*\color{scala-orange}{for}*) (i <- %*\color{scala-blue}{0}*) until N) {
          val x_x = x(i).toInt
          val y_y = y(i).toInt
          val a_a = a(i).toInt
          val b_b = b(i).toInt
          %*\color{scala-orange}{if}*) (c(i) < constraints(x_x)(y_y)(a_a)(b_b)) {
                %*\color{scala-orange}{for}*) (edge <- GetHyperedges(
                   hyperedges, Array(a_a, b_b, x_x, y_y))) {
                      hyperedgesTallies(edge) += %*\color{scala-blue}{1.0}*)
                }
                globalDistribution(GetVertex(a_a, b_b, x_x, y_y)) += %*\color{scala-blue}{1.0}*)
             }
          }
       }
       %*\color{scala-orange}{for}*) (a <- %*\color{scala-blue}{0}*) to %*\color{scala-blue}{1}*); b <- %*\color{scala-blue}{0}*) to %*\color{scala-blue}{1}*); x <- %*\color{scala-blue}{0}*) to %*\color{scala-blue}{1}*); y <- %*\color{scala-blue}{0}*) to %*\color{scala-blue}{1}*)) {
          %*\color{scala-orange}{var}*) summedAmount = %*\color{scala-blue}{0.0}*)
          %*\color{scala-orange}{val}*) associatedHyperedges = GetHyperedges(hyperedges,
             Array(a.toDouble,b.toDouble,x.toDouble,y.toDouble))
          %*\color{scala-orange}{for}*) (edgeIndex <- associatedHyperedges.indices) {
             summedAmount += hyperedgesTallies(edgeIndex)
          }
          globalDistribution(GetVertex(a,b,x,y)) = 
             globalDistribution(GetVertex(a,b,x,y)) / summedAmount
       }
       globalDistribution
    }

 }
%*\color{lgray}{.}*)
\end{lstlisting}

\subsubsection{Pyro}

Pyro's context management is integrated into Python's \verb|def| containers; or can be flexibly given implicitly, encouraging the use of stochastic functions to specify probabilistic models. Inside a container, probabilities of random variables are specified first. As Pyro is built upon PyTorch, explicit values match PyTorch types, in this case resulting in \verb|Tensor| type values.

As can be seen in Listing~\ref{fig:pyro_implementation} in the \verb|generate_global_distribution| method, Pyro's atomic sampling capabilities allow for the requirement of fewer syntactic constructs to communicate the sampling process. Each \verb|sample| call accepts a distribution, in this case either \verb|Bernoulli| or \verb|Uniform|. Upon sampling, the outcomes are tallied and normalised, before presenting the result.

\begin{lstlisting}[language=python,caption={Pyro Implementation Of EPR Simulation},captionpos=b,label={fig:pyro_implementation}]
 
 from pyro import sample
 from torch import Tensor
 from torch.autograd import Variable
 from numpy import zeros, array, fliplr, sum
 from itertools import product
 from pyro.distributions import Bernoulli, Uniform
 
 def get_vertex(a, b, x, y):
    return ((x*8)+(y*4))+(b+(a*2))
 
 def get_hyperedges(H, n):
    l = []
    for idx, e in enumerate(H):
       if n in e:
          l.append(idx)
    return l
 
 def foulis_randall_product():
    fr_edges = []
    H = [   [[[0, 0], [1, 0]], [[0, 1], [1, 1]]], 
            [[[0, 0], [1, 0]], [[0, 1], [1, 1]]]]
    for edge_a in H[0]:
       for edge_b in H[1]:
          fr_edge = []
          for vertex_a in edge_a:
             for vertex_b in edge_b:
                fr_edge.append([   
                   vertex_a[0], vertex_b[0], 
                   vertex_a[1], vertex_b[1]])
          fr_edges.append(fr_edge)
    for mc in range(0,2):
       mc_i = abs(1-mc)
       for edge in H[mc]:
          for j in range(0,2):
             fr_edge = []
             for i in range(0, len(edge)):
                edge_b = H[mc_i][i]
                vertex_a = edge[abs(i-j)]
                vertex_b = edge_b[0]
                vertex_c = edge_b[1]
                vertices_a = [
                   vertex_a[0], vertex_b[0], 
                   vertex_a[1], vertex_b[1]]
                vertices_b = [
                   vertex_a[0], vertex_c[0], 
                   vertex_a[1], vertex_c[1]]
                fr_edge.append([
                   vertices_a[mc], vertices_a[mc_i], 
                   vertices_a[mc+2], vertices_a[mc_i+2]]
                   )
                fr_edge.append([
                   vertices_b[mc], vertices_b[mc_i], 
                   vertices_b[mc+2], vertices_b[mc_i+2]])
             fr_edges.append(fr_edge)
    return fr_edges
 
 def generate_global_distribution(constraints,N):
    hyperedges = foulis_randall_product()
    hyperedges_tallies = zeros(12)
    global_distribution = zeros(16)
    while sum(global_distribution) < N:
       a = int(sample('A', Bernoulli(Variable(Tensor([0.5])))))
       b = int(sample('B', Bernoulli(Variable(Tensor([0.5])))))
       x = int(sample('X', Bernoulli(Variable(Tensor([0.5])))))
       y = int(sample('Y', Bernoulli(Variable(Tensor([0.5])))))
       value = float(sample('C', Uniform(Variable(Tensor([0.0])), 
          Variable(Tensor([1.0])))))
       if (value < constraints[x][y][a,b]):
          for edge in get_hyperedges(hyperedges, [a, b, x, y]):
             hyperedges_tallies[edge] += 1
          global_distribution[get_vertex(a, b, x, y)] += 1
    z = [0,1]
    for a, b, x, y in product(z,z,z,z):
       summed_tally = (sum(hyperedges_tallies[e] 
          for e in get_hyperedges(hyperedges, [a, b, x, y])))
       global_distribution[get_vertex(a, b, x, y)] /= summed_tally
    global_distribution *= 3
    return global_distribution
%*\color{lgray}{.}*)
\end{lstlisting}

\subsection{Input correlations for sampling}

In order to provide flexibility in investigating simulations of quantum and super-quantum correlations, 
correlations between $A$ and $B$ in the four measurement contexts are specified.
For example, the following code fragments ~\ref{fig:pyro_implementation_constraints}~\ref{fig:turing_implementation_constraints}~\ref{fig:figaro_implementation_constraints} specify that super-quantum correlations will be simulated by specifying $A$ and $B$ to be maximally correlated in three measurement contexts and maximally anti-correlated in the fourth.
With these input correlations, maximum violation of the CHSH inequalities would be expected, essentially simulating a PR box~\cite{popescu1994quantum}.

\begin{lstlisting}[language=python,caption={Implementation of input correlations in PYMC3 And Pyro},captionpos=b,label={fig:pyro_implementation_constraints}]

 constraints = [[ array([[0.5, 0], [0., 0.5]]), 
                  array([[0.5, 0], [0., 0.5]]) ],
                [ array([[0.5, 0], [0., 0.5]]), 
                  array([[0, 0.5], [0.5, 0.]]) ]]
 
 p = generate_global_distribution((constraints, 5000)
%*\color{lgray}{.}*)
\end{lstlisting}

\begin{lstlisting}[language=java,caption={Implementation of input correlations in Turing.jl},captionpos=b,label={fig:turing_implementation_constraints}]

 constraints %*\color{red}{=}*) [[ [[0.5, 0.0], [0.0, 0.5]], 
                  [[0.5, 0.0], [0.0, 0.5]] ], 
                [ [[0.5, 0.0], [0.0, 0.5]], 
                  [[0.0, 0.5], [0.5, 0.0]] ]]
 
 p %*\color{red}{=}*) %*\color{darkolivegreen}{generate\_global\_distribution}*)(constraints, 5000)
%*\color{lgray}{.}*)
\end{lstlisting}

\begin{lstlisting}[language=java,caption={Implementation of input correlations in Figaro},captionpos=b,label={fig:figaro_implementation_constraints}]
 
 %*\color{scala-orange}{val}*) constraints = Array(
    Array( Array(Array(%*\color{scala-blue}{0.5}*), %*\color{scala-blue}{0.0}*)), Array(%*\color{scala-blue}{0.0}*), %*\color{scala-blue}{0.5}*))), 
           Array(Array(%*\color{scala-blue}{0.5}*), %*\color{scala-blue}{0.0}*)), Array(%*\color{scala-blue}{0.0}*), %*\color{scala-blue}{0.5}*)))),
    Array( Array(Array(%*\color{scala-blue}{0.5}*), %*\color{scala-blue}{0.0}*)), Array(%*\color{scala-blue}{0.0}*), %*\color{scala-blue}{0.5}*))), 
           Array(Array(%*\color{scala-blue}{0.0}*), %*\color{scala-blue}{0.5}*)), Array(%*\color{scala-blue}{0.5}*), %*\color{scala-blue}{0.0}*)))))

 %*\color{scala-orange}{val}*) P = GenerateGlobalDistribution(constraints,%*\color{scala-blue}{5000}*))
%*\color{lgray}{.}*)
\end{lstlisting}

\subsection{Specifying the CHSH inequalities}
For each of the four PPLs, code is specified ~\ref{fig:pymc_pyro_chsh}~\ref{fig:turing_chsh}~\ref{fig:figaro_chsh} that implements the system of four CHSH inequalities.
The outcome is a vector of four Boolean values expressing whether the respective inequality was violated. 

\subsubsection{PYMC3 And Pyro}

\begin{lstlisting}[language=python,caption={Specification of the CHSH inequalities in PYMC3 And Pyro},captionpos=b,label={fig:pymc_pyro_chsh}]

 def equality(v1,v2,v3,v4):
    def f1(v1,v2):
       return abs((2 * (p[v1] + p[v2])) - 1)
    def f2(v1,v2,v3,v4):
       return (p[v1] + p[v2]) - (p[v3] + p[v4])
    delta = 0.5 * ( 
       (f1(0,1) - f1(4,5)) + (f1(8,9) - f1(12,13)) + 
       (f1(0,2) - f1(4,6)) + (f1(8,10) - f1(12,14)))
    return 2 * (1 + delta) >= abs(
       (v1*f2(0,3,1,2)) + (v2*f2(4,7,5,6)) +
       (v3*f2(8,11,9,10)) + (v4*f2(12,15,13,14)))
 
 tests = [
    equality(1,1,1,-1),
    equality(1,1,-1,1),
    equality(1,-1,1,1),
    equality(-1,1,1,1)]
%*\color{lgray}{.}*)
\end{lstlisting}

\subsubsection{Turing.jl}

\begin{lstlisting}[language=java,caption={Specification of the CHSH inequalities in Turing.jl},captionpos=b,label={fig:turing_chsh}]
 
 %*\color{red}{function}*) %*\color{darkolivegreen}{equality}*)(v1,v2,v3,v4)
    %*\color{red}{function}*) %*\color{darkolivegreen}{f1}*)(v1,v2)
       %*\color{darkolivegreen}{abs}*)((2 %*\color{red}{\ding{83}}*) (p[v1] %*\color{red}{+}*) p[v2])) %*\color{red}{-}*) 1)
    %*\color{red}{end}*)
    %*\color{red}{function}*) %*\color{darkolivegreen}{f2}*)(v1,v2,v3,v4)
       (p[v1] %*\color{red}{+}*) p[v2]) %*\color{red}{-}*) (p[v3] %*\color{red}{+}*) p[v4])
    %*\color{red}{end}*)
    delta %*\color{red}{=}*) 0.5 %*\color{red}{\ding{83}}*) ( 
       (%*\color{darkolivegreen}{f1}*)(1,2) %*\color{red}{-}*) %*\color{darkolivegreen}{f1}*)(5,6)) %*\color{red}{+}*) (%*\color{darkolivegreen}{f1}*)(9,10) %*\color{red}{-}*) %*\color{darkolivegreen}{f1}*)(13,14)) %*\color{red}{+}*) 
       (%*\color{darkolivegreen}{f1}*)(1,3) %*\color{red}{-}*) %*\color{darkolivegreen}{f1}*)(5,7)) %*\color{red}{+}*) (%*\color{darkolivegreen}{f1}*)(9,11) %*\color{red}{-}*) %*\color{darkolivegreen}{f1}*)(13,15)))
    (2 %*\color{red}{\ding{83}}*) (1 %*\color{red}{+}*) delta)) >= %*\color{darkolivegreen}{abs}*)(
       (v1 %*\color{red}{\ding{83}}*) %*\color{darkolivegreen}{f2}*)(1,4,2,3)) %*\color{red}{+}*) (v2 %*\color{red}{\ding{83}}*) %*\color{darkolivegreen}{f2}*)(5,8,6,7)) %*\color{red}{+}*)
       (v3 %*\color{red}{\ding{83}}*) %*\color{darkolivegreen}{f2}*)(9,12,10,11)) %*\color{red}{+}*) (v4 %*\color{red}{\ding{83}}*) %*\color{darkolivegreen}{f2}*)(13,16,14,15)))
 %*\color{red}{end}*)

 tests = [
    %*\color{darkolivegreen}{equality}*)(1,1,1,-1),
    %*\color{darkolivegreen}{equality}*)(1,1,-1,1),
    %*\color{darkolivegreen}{equality}*)(1,-1,1,1),
    %*\color{darkolivegreen}{equality}*)(-1,1,1,1)]
%*\color{lgray}{.}*)
\end{lstlisting}

\subsubsection{Figaro}

\begin{lstlisting}[language=java,caption={Specification of the CHSH inequalities in Figaro},captionpos=b,label={fig:figaro_chsh}]

 %*\color{scala-orange}{def}*) %*\color{scala-yellow}{Equality}*)(v1: %*\color{scala-orange}{Int}*)%*\color{scala-orange}{,}*) v2: %*\color{scala-orange}{Int}*)%*\color{scala-orange}{,}*) v3: %*\color{scala-orange}{Int}*)%*\color{scala-orange}{,}*) v4: %*\color{scala-orange}{Int}*)): %*\color{scala-orange}{Boolean}*) = {
    %*\color{scala-orange}{def}*) %*\color{scala-yellow}{f1}*)(v1: %*\color{scala-orange}{Int}*)%*\color{scala-orange}{,}*) v2: %*\color{scala-orange}{Int}*)): %*\color{scala-orange}{Double}*) = {
       Math.abs((%*\color{scala-blue}{2}*) * (p(v1) + p(v2))) - %*\color{scala-blue}{1}*))
    }
    %*\color{scala-orange}{def}*) %*\color{scala-yellow}{f2}*)(v1: %*\color{scala-orange}{Int}*)%*\color{scala-orange}{,}*) v2: %*\color{scala-orange}{Int}*)%*\color{scala-orange}{,}*) v3: %*\color{scala-orange}{Int}*)%*\color{scala-orange}{,}*) v4: %*\color{scala-orange}{Int}*)): %*\color{scala-orange}{Double}*) = {
       (p(v1) + p(v2)) - (p(v3) + p(v4))
    }
    %*\color{scala-orange}{val}*) delta = %*\color{scala-blue}{0.5}*) * (
       (f1(%*\color{scala-blue}{0}*)%*\color{scala-orange}{,}*)%*\color{scala-blue}{1}*)) - f1(%*\color{scala-blue}{4}*)%*\color{scala-orange}{,}*)%*\color{scala-blue}{5}*))) + (f1(%*\color{scala-blue}{8}*)%*\color{scala-orange}{,}*)%*\color{scala-blue}{9}*)) - f1(%*\color{scala-blue}{12}*)%*\color{scala-orange}{,}*)%*\color{scala-blue}{13}*))) +
       (f1(%*\color{scala-blue}{0}*)%*\color{scala-orange}{,}*)%*\color{scala-blue}{2}*)) - f1(%*\color{scala-blue}{4}*)%*\color{scala-orange}{,}*)%*\color{scala-blue}{6}*))) + (f1(%*\color{scala-blue}{8}*)%*\color{scala-orange}{,}*)%*\color{scala-blue}{10}*)) - f1(%*\color{scala-blue}{12}*)%*\color{scala-orange}{,}*)%*\color{scala-blue}{14}*))))
    (%*\color{scala-blue}{2}*) * (%*\color{scala-blue}{1}*) + delta)) >= Math.abs(
       (v1*f2(%*\color{scala-blue}{0}*)%*\color{scala-orange}{,}*)%*\color{scala-blue}{3}*)%*\color{scala-orange}{,}*)%*\color{scala-blue}{1}*)%*\color{scala-orange}{,}*)%*\color{scala-blue}{2}*))) + (v2*f2(%*\color{scala-blue}{4}*)%*\color{scala-orange}{,}*)%*\color{scala-blue}{7}*)%*\color{scala-orange}{,}*)%*\color{scala-blue}{5}*)%*\color{scala-orange}{,}*)%*\color{scala-blue}{6}*))) +
       (v3*f2(%*\color{scala-blue}{8}*)%*\color{scala-orange}{,}*)%*\color{scala-blue}{11}*)%*\color{scala-orange}{,}*)%*\color{scala-blue}{9}*)%*\color{scala-orange}{,}*)%*\color{scala-blue}{10}*))) + (v4*f2(%*\color{scala-blue}{12}*)%*\color{scala-orange}{,}*)%*\color{scala-blue}{15}*)%*\color{scala-orange}{,}*)%*\color{scala-blue}{13}*)%*\color{scala-orange}{,}*)%*\color{scala-blue}{14}*))))
 }
 
 %*\color{scala-orange}{val}*) tests = Array[%*\color{scala-orange}{Boolean}*)](
    Equality(%*\color{scala-blue}{1}*)%*\color{scala-orange}{,}*)%*\color{scala-blue}{1}*)%*\color{scala-orange}{,}*)%*\color{scala-blue}{1}*)%*\color{scala-orange}{,}*)%*\color{scala-blue}{-1}*))%*\color{scala-orange}{,}*)
    Equality(%*\color{scala-blue}{1}*)%*\color{scala-orange}{,}*)%*\color{scala-blue}{1}*)%*\color{scala-orange}{,}*)%*\color{scala-blue}{-1}*)%*\color{scala-orange}{,}*)%*\color{scala-blue}{1}*))%*\color{scala-orange}{,}*)
    Equality(%*\color{scala-blue}{1}*)%*\color{scala-orange}{,}*)%*\color{scala-blue}{-1}*)%*\color{scala-orange}{,}*)%*\color{scala-blue}{1}*)%*\color{scala-orange}{,}*)%*\color{scala-blue}{1}*))%*\color{scala-orange}{,}*)
    Equality(%*\color{scala-blue}{-1}*)%*\color{scala-orange}{,}*)%*\color{scala-blue}{1}*)%*\color{scala-orange}{,}*)%*\color{scala-blue}{1}*)%*\color{scala-orange}{,}*)%*\color{scala-blue}{1}*))
 )
%*\color{lgray}{.}*)
\end{lstlisting}

Upon conducting simulations using the input correlations given above, the predicted maximum violation of the CHSH inequalities were observed for all four PPLs specified above.

\section{Numerical Evaluation}
We conducted several experiments to compare the numerical accuracy and execution time of the different implementations.

\subsection{Accuracy Of Tests}

For all PPLs, statistical outputs confirming the success of the FR product (and consequently the no-signalling condition) are given with an acceptable margin of error. While this is consequent of more than a single factor, it is perceived that the largest contributor to accuracy is the computation of random values for each PPL. What can be observed is that, with larger sample sizes (bearing more perfectly random distributions), that the margin of error decreases, as can be seen below. This is typical, as the experiment design's normalisation process is dependent on the even spread of tallies across the global distribution, and more specifically, the degree to which the sampler is random. It should also be considered that the various PPLs apply data types that round values for a loss of statistical precision where it may serve meaning. For example, while a single value may lose a minute portion of its whole beyond the decimal point (due to automatic rounding), when calculating a handful of these values per iteration of some few thousand iterations, the difference becomes observable.

\begin{figure}[h]
\captionsetup{justification=centering}
\centering
\includegraphics[width=13cm]{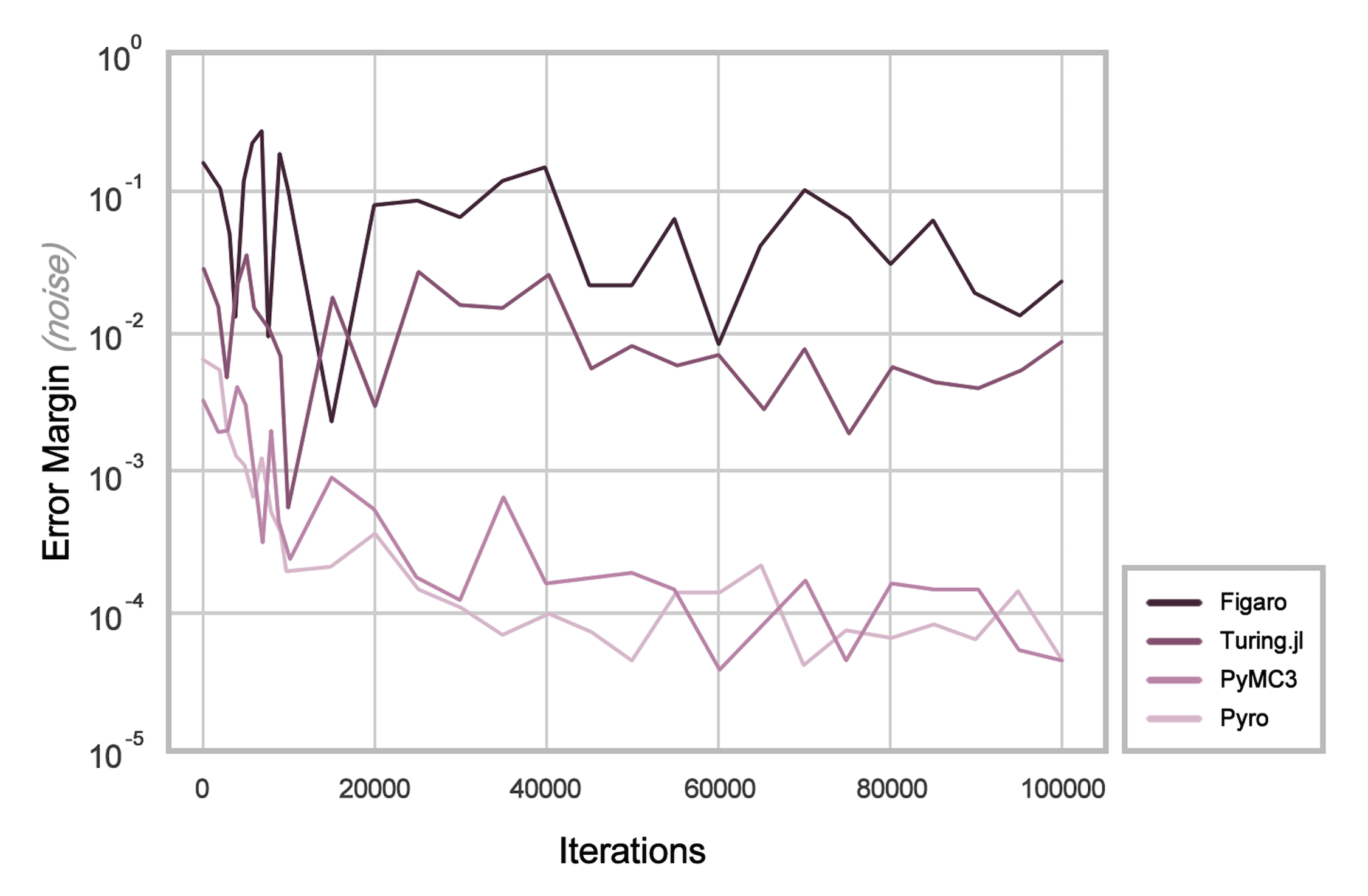}
\caption{Accuracy Of EPR Experimentation}\label{fig:tests_accuracy}
\end{figure}

From observing the results in Fig.~\ref{fig:tests_accuracy}, it can be seen immediately that before the first 20,000 iterations, all PPLs exhibit substantial noise that rules out the possibility of accounting for said window of results. In Monte Carlo inference, noise of this kind is common, where accounting for 'burn-in' iterations at the beginning of the sampling process may possibly minimize the unpredictability of the results. It is also seen that Figaro and Turing.jl have consistently greater margins of error than those of PyMC3 and Pyro. Overcoming this error would be achieved through improved float precision where applicable in either PPL's programming language. It cannot be said which of either PyMC3 or Pyro display the most accurate results. Of interest, it is perceived that where other PPLs may have required multiple instances of the entire sampling process to tally the global distribution, Pyro could accurately equate the global distribution atomically, thus improving its accuracy.

\subsection{Elapsed Time Of Execution}

Another statistic that has been observed is the compilation time of each PPL, which typically increases with number of iterations. For relativity of results, it should be noted that all non-accelerated tests of this kind were executed within a bash execution terminal, on a Macintosh operating system, bearing a 45nm ``Penryn" 2.4 GHz Intel ``Core 2 Duo" processor, and 4 GB of SDRAM, whereas all accelerated tests were executed within a bash execution terminal of Amazon Web Services Linux (2nd distribution). The specification of the 'Elastic-Compute Cloud' on which the Linux distribution executed was a 2018 ``p2.xlarge" 2.7 GHz Intel Broadwell processor, with 61 GB of SDRAM. The instance also provides an NVIDIA GK210 GPU multi-vCPU (count of 4) processor, and 12 GB of GPU RAM.

\begin{figure}[h]
\captionsetup{justification=centering}
\centering
\includegraphics[width=13cm]{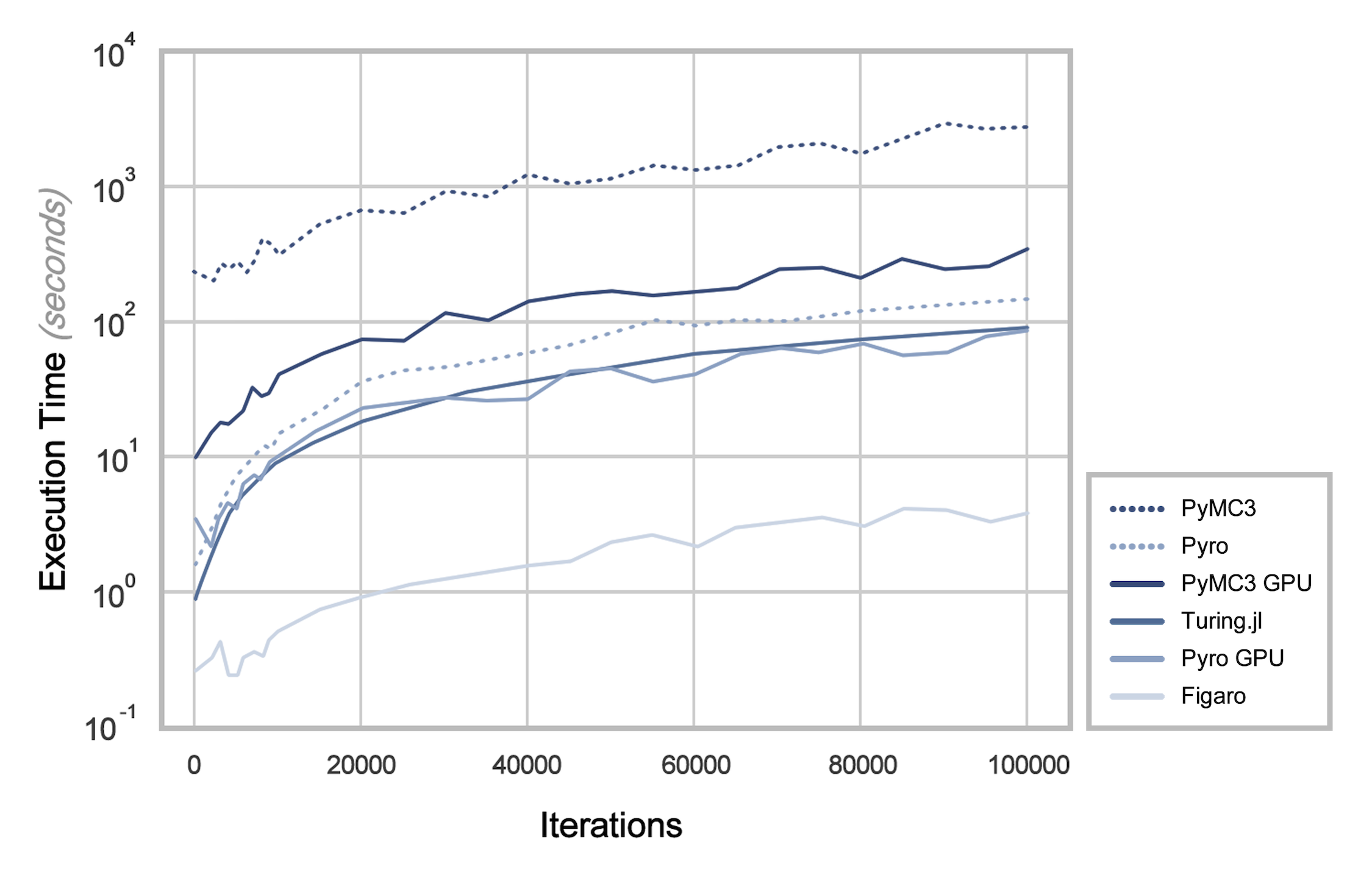}
\caption{Execution time of EPR Experimentation}\label{fig:tests_time}
\end{figure}

In Fig.~\ref{fig:tests_time}, it can be seen that among the accelerated results, the fastest PPL is Figaro by scale of almost an entire logarithmic unit. Thereafter, Pyro and Turing.jl tie in second position, however Pyro demonstrates inference of a stabler nature. Despite the Theano architecture's utilisation of the supplied GPU, PyMC3 is then the most affected by size of experimental setup. When the Theano architecture is non-accelerated, it can be seen that PyMC3's performance drastically decreases. For comparison, Pyro has also been tested on a non-accelerated architecture, where the difference in performance is reasonably smaller than that of PyMC3. This forms the suggestion that PyMC3 should not be applied in non-accelerated environments. In all cases, it can be seen that all PPLs exhibit a linear order of growth in the given scenario.

\section{Discussion}

Recall that the challenge posed was how to develop probabilistic program which can simulate quantum correlations in an EPR experiment.
The solution adopted was to program a hypergraph formalism to underpin the simulations.
This formalism is modular where the FR product of the modules is used to impose the no-signalling constraint.
In execution, all four PPLs successfully simulated an EPR experiment producing quantum correlations. 
Therefore, we conclude that the hypergraph formalism has been shown to be a promising basis for such simulations.
In addition, the hypergraph formalism is also rendered into program syntax in a fairly straightforward way.
However, the  formalism does pose a challenge with respect to the accessibility criterion of the PPLs.
The challenge is due to an inherent ambiguity present in the composite hypergraph produced by the FR product, which has 12 edges in the EPR experiment. Four of those edges represent ``actual" measurement contexts (depicted in Figure \ref{fig:p16}), whilst the remaining eight edges impose the no-signalling condition. The hypergraph formalism is agnostic to this distinction, which is important to distinguish when designing and programming simulations.

On the other hand, the strength of the hypergraph formalism is its flexibility and modularity. In particular, modularity offers the potential to cover a wide variety of experimental designs whilst at the same time offering a conceptually simple route to program specification. We have shown that the  EPR experiment is based on two modules which represent measurements on the individual systems $A$ and $B$. 
More generally, joint measurements on multiple systems, and the constraints they must satisfy can then be expressed in terms of a composition operator that combines the modules into a suitable global data structure in the program, which underpins both the sampling and simulation.

With regard to sampling, an immediate suggestion is stronger control-flow integration in the sampling process. Rather than repeatedly generating distributions that are  indexed for random results, all PPLs should offer atomic sampling similar to the likes of Pyro or Figaro, where single values could be observed, or returned from a \verb|Stream| primitive in a single procedure. As a sampling process contains variables that are akin to those of iterative loops, it may also serve PPLs to re-imagine the sampling process as a paradigm of the contained programming language, rather than as a single procedure that operates in isolation of the entire program. Providing control over each iteration of the distribution could also improve the legibility of the program, while minimising convolution in the procedures that typically come afterwards.

In terms of the qualitative comparison between the four PPLs, Figaro demonstrated the most benefits for general usage. This could be seen in its capacity to deliver specialised features where other PPLs could not. For what features the alternatives provide, they may appropriately match Figaro i.e., consider that PyMC3 offers more configurations for distributions described in probabilistic models, or that Pyro achieves control-flow independence with fewer syntactic constructs. Such arguments have been overlooked when taking into account the efforts that the main developer, Charles River Analytics, has made to ensure that its PPL is competitively implemented in wider applications. For the likes of accessibility, Figaro's origins in Scala do not present the same benefits as Pyro in agile development. However, the simulation of quantum correlations is not a rigorous process. Furthermore, acceleration of PPLs was observed to be minimal in difference (exempt of PyMC3) for the case of the experimentation. Thus it wouldn't be perceived that this is a determinant factor.

In experimentation, we found that Pyro provided the syntactic constructs needed to neatly describe its processes in fewer procedures than those of the others. While PyMC3's origins in Python also made it an expressive alternative, the excessive nomenclature surrounding the declarations of methods and data-types for both Turing.jl and Figaro convoluted their descriptions. While in comparison to the other PPLs, Figaro's accuracy is inferior, it could be argued that the sample iterations describe an experimental setting that does not consider improving float precision. Coupled with the trend of Figaro's improvement in its number of iterations, and the measure of accuracy between PPLs may converge. The same cannot be said for the time complexity of the EPR experimentation, where it was observed that PyMC3's compilation grew substantially with the number of sample iterations being executed. Still, in instances where accuracy is a key factor and limitations are perceived in Pyro's functionality, PyMC3 would be the suitable alternative.

\section{Conclusion}
Probabilistic programming offers new possibilities for quantum physicists to specify and simulate experiments, such as the EPR experiment illustrated in this article. 
This is particularly relevant for experiments requiring advanced statistical inference on unknown parameters, especially in the case of techniques that involve large amounts of data and computational time.
Furthermore, probabilistic machine learning models that are conveniently expressed in probabilistic programming languages can advance our understanding of the underlying physics of the experiments.

It is important to note that the benefits of  probabilistic programming are not restricted to experiments involving the analysis of quantum correlations.
Since any probabilistic programming language is based on random variables, we can ask the question what exactly is a random variable in quantum physics.
Focusing on a single measurement context, due to the normalization constraint, we can think of the measurement context as a (conditional) probability distribution over random variables, which describes the measurement outcomes.
The probability distribution is a normalized measure over the sigma algebra defined by the outcomes.
This measure is defined via Born's rule, that is, the quantum state is embedded in the measure.
An EPR experiment is essentially a state preparation protocol where deterministic operations are embedded in the quantum state (the unitary operations leading to its preparation), followed by the measurement, which results in stochastic outcomes.
More generally, we can think of a larger system where we only measure a subsystem.
This leads to quantum channels, which are described by completely positive trace preserving maps.
A quantum channel, however, must be deterministic in the sequence of events, and, for instance, a measurement choice at a later time step cannot depend on the outcome of a previous measurement.
We must factor in such classical and quantum memory effects, as well as the potentially indeterminate causal order of events.
The quantum comb~\cite{chiribella2008quantum,chiribella2009theoretical} or process matrix~\cite{oreshkov2012quantum,pollock2018complete,pollock2018operational} formalism addresses these more generic requirements.
Either formalism introduces a generalized Born's rule, where deterministic and stochastic parts of the system clearly separate, and thus give a clear way of defining random variables.
Probabilistic programming offers potential in expressing models designed in these frameworks.

\section{Acknowledgements}
This research was supported by the Asian Office of Aerospace Research and Development (AOARD) grant: FA2386-17-1-4016. This research was supported by Perimeter Institute for Theoretical Physics. Research at Perimeter Institute is supported by the Government of Canada through Industry Canada and by the Province of Ontario through the Ministry of Economic Development and Innovation.

\end{document}